\documentclass{article}
\usepackage{graphicx}
\usepackage{amsmath}
\usepackage[version=4]{mhchem}
\usepackage{bm}
\usepackage{setspace}
\usepackage{multirow}
\usepackage{colortbl}

\newcommand{\PreserveBackslash}[1]{\let\temp=\\#1\let\\=\temp}
\newcolumntype{C}[1]{>{\PreserveBackslash\centering}p{#1}}
\newcolumntype{R}[1]{>{\PreserveBackslash\raggedleft}p{#1}}
\newcolumntype{L}[1]{>{\PreserveBackslash\raggedright}p{#1}}

\usepackage[numbers]{natbib}

\usepackage[main, final]{neurips_2026}

\usepackage[utf8]{inputenc} \usepackage[T1]{fontenc}    \usepackage{hyperref}       
\usepackage{url}            \usepackage{booktabs}       \usepackage{amsfonts}       \usepackage{nicefrac}       \usepackage{microtype}      \usepackage{xcolor}         \usepackage{enumitem}       

\title{
  CrystalREPA: Transferring Physical Priors from Universal MLIPs to Crystal Generative Models
  }

\author{
  Chengqian Zhang{$^{1,2}$},~
  Yucheng Jin{$^{1}$},~
  Duo Zhang{$^{1,2}$},~
  Tiejun Li{$^{2,3,4,\ast}$},~
  Han Wang{$^{5,6,}$}\thanks{
    Correspondence to:
    \nolinkurl{tieli@pku.edu.cn},
    \nolinkurl{wang_han@iapcm.ac.cn}
  }\\
\vspace{-1pt}{\small~}\\
{$^{1}$}AI for Science Institute, Beijing, China;\\
  {$^{2}$}Center for Data Science, Peking University, Beijing, China;\\
  {$^{3}$}LMAM and School of Mathematical Sciences, Peking University, Beijing, China;\\
  {$^{4}$}Center for Machine Learning Research, Peking University, Beijing, China;\\
  {$^{5}$}National Key Laboratory of Computational Physics,\\
  Institute of Applied Physics and Computational Mathematics, Beijing, China;\\
  {$^{6}$}HEDPS, CAPT, College of Engineering, Peking University, Beijing, China;
}

\begin{document}

\maketitle

\begin{abstract}
Crystal generative models mainly learn what stable crystals look like, with little explicit supervision for what makes them stable.
We reveal a substantial representation gap between state-of-the-art crystal generative models and pretrained universal machine learning interatomic potentials (MLIPs) via energy probing, and show this gap can be closed by a simple training-time alignment.
We propose Crystal REPresentation Alignment (CrystalREPA), a plug-and-play framework that aligns the atom-wise hidden states of generative encoders with frozen MLIP representations through an element-aware contrastive objective, transferring stability-aware atomistic priors with marginal training overhead and no additional inference cost.
Across three generative frameworks, ten MLIP teachers, and two benchmark datasets, CrystalREPA consistently improves the thermodynamic stability, structural validity, and structural fidelity of generated crystals.
Equally important, we find that an MLIP's transfer effectiveness is poorly predicted by its accuracy on standard leaderboards (e.g., Matbench Discovery) but strongly predicted by the distinguishability of its atom-wise representation space, yielding a practical, accuracy-independent criterion for selecting MLIP teachers for generative transfer.
Code is available at \url{https://github.com/Chengqian-Zhang/CrystalREPA}.
\end{abstract}

\begin{figure}[h]
  \centering
  \includegraphics[width=0.98\textwidth]{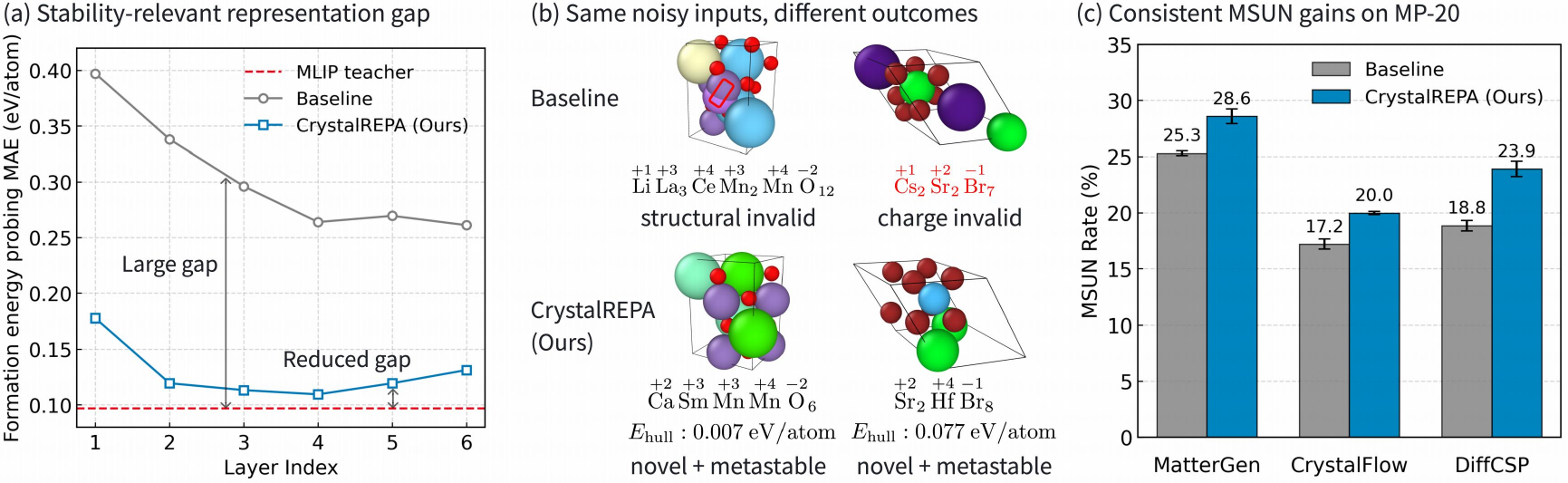}
  \caption{
  CrystalREPA transfers stability-aware priors from universal MLIPs to crystal generative models.
  (a) Formation energy probing reveals that vanilla generative representations are substantially less predictive of stability than pretrained MLIP representations, while CrystalREPA narrows this gap through representation alignment.
  (b) Given the same noisy initial structures, vanilla CrystalFlow can generate physically invalid crystals, whereas CrystalREPA generates novel and metastable structures.
(c) CrystalREPA also yields consistent gains in MSUN (Metastable, Unique, and Novel) rate across MatterGen, CrystalFlow, and DiffCSP, indicating better discovery-oriented generation quality.
  Panels (a) and (b) use CrystalFlow, and panel (c) compares all three generators; all CrystalREPA results shown here use DPA-3.1-3M as the MLIP teacher.
}
  \label{fig:summary}
\end{figure}

\section{Introduction}
The inverse design of crystalline materials aims to generate novel structures with desired stability and functionality, but remains challenging due to the vast combinatorial space of atomic configurations and the stringent periodic constraints of crystals~\cite{cheng2026artificial}. 
Recent advances in generative modeling, particularly diffusion- and flow-based approaches, have enabled direct sampling of candidate crystal structures from data-driven distributions, marking a significant step toward automated materials discovery~\cite{metni2026generative, zeni2025generative, luo2025crystalflow, jiao2023crystal}.

Despite this progress, current crystal generative models mainly learn what stable crystals look like, rather than what makes a configuration stable.
Popular benchmark datasets such as MP-20~\cite{xie2022crystal} and Alex-MP-20~\cite{zeni2025generative} consist primarily of stable or metastable equilibrium structures.
Training on such data encourages models to reproduce plausible geometries and compositions, but exposes them to only a narrow region of the potential energy landscape.
In particular, the model rarely observes non-equilibrium, distorted, or high-energy configurations, nor does it receive explicit energy or force labels that quantify how unfavorable such configurations are.
As a result, generated crystals may appear geometrically plausible while still deviating from low-energy or physically reliable regions of configuration space.

Universal machine learning interatomic potentials (MLIPs) offer a complementary source of supervision~\cite{yuan2026foundation}.
Unlike crystal generative models, MLIPs are trained on large-scale atomistic datasets with explicit energy and force labels, covering not only relaxed structures but also perturbed, non-equilibrium, and high-energy configurations.
This supervision encourages MLIPs to learn representations that are sensitive to variations in the potential energy surface and therefore relevant to stability.
Empirically, this supervision difference manifests as a substantial \emph{representation gap}: features learned by crystal generative models are far less predictive of formation energy than those of pretrained universal MLIPs.
This motivates the central question of this work: \textit{can we transfer stability-aware atomistic priors from universal MLIPs to crystal generative models without compromising generative flexibility?}

To address this question, we propose \textit{Crystal REPresentation Alignment (CrystalREPA)}, a simple and general framework for transferring stability-aware priors from pretrained universal MLIPs to crystal generative models.
Inspired by representation alignment in image generation~\cite{yu2025representation}, CrystalREPA aligns the atom-wise representations of generative encoders with those of pretrained universal MLIPs through an element-aware contrastive objective, encouraging the crystal generative model to learn features predictive of stability.
We validate CrystalREPA across multiple popular generative frameworks, universal MLIPs, and benchmark datasets.
Extensive experiments demonstrate that CrystalREPA consistently improves the thermodynamic stability, structural validity, and structural fidelity of generated crystals while preserving diversity, introducing only marginal training overhead and no extra inference cost.
Beyond performance gains, we further show that the representational distinguishability of the teacher MLIPs strongly correlates with downstream generative quality, suggesting it as an important factor associated with transfer effectiveness.

Our contributions are summarized as follows:
\begin{itemize}[leftmargin=1.2em, itemsep=1pt, topsep=2pt, parsep=0pt]
  \item We make the geometry-vs-energy gap quantitative: via formation-energy probing, crystal generative model representations are far less predictive of stability than pretrained universal MLIPs.
  \item We propose CrystalREPA, which transfers stability-aware priors by aligning atom-wise hidden states with frozen MLIP representations via an element-aware contrastive objective, and consistently improves stability, validity, and fidelity across three generative frameworks, ten MLIP teachers, and two datasets, at marginal training cost and no inference cost.
  \item We find that an MLIP teacher's transfer effectiveness is predicted by its atom-wise representation distinguishability rather than leaderboard accuracy, yielding a practical teacher-selection criterion.
\end{itemize}

\section{Related Work}

\textbf{Crystal Generative Models.}
Generative modeling has become a promising paradigm for inverse crystal design, with diffusion-based methods~\cite{ho2020denoising, song2021scorebased} becoming the dominant approach for modeling periodic atomic structures~\cite{xie2022crystal, luo2023towards, jiao2023crystal, zeni2025generative, cornet2025kinetic, lin2024equicsp}.
Representative models such as DiffCSP~\cite{jiao2023crystal} and MatterGen~\cite{zeni2025generative} adopted joint diffusion frameworks that simultaneously model atomic composition, fractional coordinates, and lattice parameters, enabling end-to-end sampling of candidate materials from data distributions.
Subsequent studies have incorporated crystallographic priors such as symmetry constraints to reduce the effective search space and improve structural validity~\cite{jiao2024space, levy2025symmcd, kelvinius2025wyckoffdiff, chang2025space}.
In parallel, flow-based approaches~\cite{lipman2023flow, albergo2023building, liu2023flow} have recently been introduced for crystal generation~\cite{miller2024flowmm, sriram2024flowllm, luo2025crystalflow}. 
By learning continuous transport from a simple base distribution to the data distribution, these methods often offer more efficient sampling while retaining competitive generation quality.
Despite these advances, most crystal generative models learn the geometry of stable crystals without supervision from the potential energy surface that governs stability.

\textbf{Universal MLIPs.}
Universal MLIPs have recently achieved robust generalization across broad chemical spaces by training on large-scale atomistic datasets such as MPtrj~\cite{deng2023chgnet},  Alexandria~\cite{schmidt2023machine, schmidt2024improving}, and OMat24~\cite{barroso2024open} with explicit energy and force supervision~\cite{riebesell2025framework, mazitov2025pet, fu2025learning, lee2025flashtp, zhou2026matris, kim2026optimizing, tan2026high, lysogorskiy2026graph, rhodes2025orbv3atomisticsimulationscale, bochkarev2024graph,zhang2025graphneuralnetworkera, batatia2025foundation, yang2024mattersim,park2024scalable, neumann2024orbfastscalableneural,zhang2024dpa,wood2025uma}.
Beyond their use for accurate energy prediction and molecular dynamics, pretrained MLIPs have been shown to encode atomistic representations sensitive to variations in the potential energy surface~\cite{zhang2024dpa, zhang2026multi}.
However, how to exploit these pretrained representations as transferable stability-aware priors for crystal generation remains largely unexplored.

\textbf{Position of our work.} 
Our work connects crystal generative modeling with atomistic representation learning.
Instead of improving generators through architectural design, symmetry constraints, or sampling procedures, we enhance their internal representations with stability-aware priors.
Instead of using MLIPs for energy prediction or post-generation relaxation, we use pretrained universal MLIPs as frozen representation teachers during generative training.
Building on representation alignment in image generation~\cite{yu2025representation}, our contribution lies in transferring energy- and force-supervised MLIP representations to crystal generative models through atom-wise, element-aware alignment.

\section{Preliminaries}

Our method builds on two components: a crystal generative model that produces atom-wise representations during denoising or transport, and a pretrained universal MLIP that provides atom-wise representations supervised by energy and forces.
We use CrystalFlow~\cite{luo2025crystalflow} as an example to introduce the former, and then review how universal MLIPs learn representations from energy and force supervision.

\subsection{Crystal generative modeling}

A crystal with $N$ atoms in the unit cell is represented by $\mathcal{M} = (\mathbf{A}, \mathbf{F}, \mathbf{L})$, where $\mathbf{A} = [\mathbf{a}_1^\top; \mathbf{a}_2^\top; \dots; \mathbf{a}_N^\top] \in \mathbb{R}^{N \times a}$ denotes atom types encoded as categorical vectors.
$\mathbf{F} =[\mathbf{f}_1^\top; \mathbf{f}_2^\top; \dots; \mathbf{f}_N^\top] \in [0,1)^{N \times 3}$ denotes fractional coordinates, and $\mathbf{L} \in \mathbb{R}^{6}$ denotes a rotation-invariant lattice parameterization derived from the lattice matrix $\tilde{\mathbf{L}} = [\mathbf{l}_1^\top; \mathbf{l}_2^\top;\mathbf{l}_3^\top] \in \mathbb{R}^{3 \times 3}$.
Following CrystalFlow, crystal generation is formulated as learning a continuous transport map from a simple prior distribution to the data distribution.

Specifically, let $\mathcal{M}_1 = (\mathbf{A}_1, \mathbf{F}_1, \mathbf{L}_1)$ be a crystal sampled from the data distribution and let $\mathcal{M}_0 = (\mathbf{A}_0, \mathbf{F}_0, \mathbf{L}_0)$ be sampled from the corresponding priors.
A continuous path $\mathcal{M}_t = (\mathbf{A}_t, \mathbf{F}_t, \mathbf{L}_t)$, with $t \in [0,1]$, interpolates between noise and data, where $t=0$ corresponds to the prior and $t = 1$ corresponds to the clean crystal.
This path induces target velocity fields $\mathbf{u}_t^A$, $\mathbf{u}_t^F$, and $\mathbf{u}_t^L$ for atom types, fractional coordinates, and lattices, respectively.
The mathematical formulation of this path is provided in Appendix~\ref{sec:crystalflow}.

CrystalFlow learns three time-dependent vector fields, $\mathbf{v}_\theta^A(\mathcal{M}_t, t)$,  $\mathbf{v}_\theta^F(\mathcal{M}_t, t)$, and $\mathbf{v}_\theta^L(\mathcal{M}_t, t)$, to regress these targets.
These vector fields are parameterized by a periodic $E(3)$-equivariant graph neural network with $K$ message-passing layers, whose final atom-wise hidden states are denoted by $f^{(K)}(\mathcal{M}_t, t) \in \mathbb{R}^{N \times d_1}$.
The velocity fields are predicted as
\begin{equation}
\label{eq:predict}
    \mathbf{v}_\theta^A = \varphi_{A}(f^{(K)}(\mathcal{M}_t, t)), \quad
    \mathbf{v}_\theta^F = \varphi_{F}(f^{(K)}(\mathcal{M}_t, t)), \quad
    \mathbf{v}_\theta^L = \varphi_{L}(\frac{1}{N} \sum_i^N f_i^{(K)}(\mathcal{M}_t, t)),
\end{equation}
where $\varphi_A$, $\varphi_F$ and $\varphi_L$ are prediction heads.
The model is trained with a weighted regression objective
\begin{equation}
\label{eq:loss_gen}
  \mathcal{L}_{\mathrm{gen}} = 
  \lambda_A\mathcal{L}_A(\mathbf{v}_\theta^A, \mathbf{u}_t^A) + 
  \lambda_F\mathcal{L}_F(\mathbf{v}_\theta^F, \mathbf{u}_t^F) + 
  \lambda_L\mathcal{L}_L(\mathbf{v}_\theta^L, \mathbf{u}_t^L),
\end{equation}
where $\mathcal{L}_A$, $\mathcal{L}_F$, and $\mathcal{L}_L$ are the corresponding regression losses for atom types, fractional coordinates, and lattice variables.

For our purpose, the key observation is that the generative encoder produces atom-wise representations on corrupted intermediate structure $\mathcal{M}_t$.
These representations are optimized for denoising or transport, but are not supervised by signals from the potential energy surface.

\subsection{Universal MLIPs}
Universal MLIPs are atomistic models trained on large-scale crystal datasets with explicit energy and force supervision.
Given a crystal structure $\mathcal{M} = (\mathbf{A}, \mathbf{F}, \mathbf{L})$ with energy label $E$ and force label $\mathbf{F}_{\mathrm{force}}$, an MLIP produces atom-wise representations through an atomistic encoder $\tilde{f}$ and predicts the total energy by aggregating atom-level contributions:
\begin{equation}
\label{eq:mlip}
    \hat{E} = \sum_{i=1}^N \varphi_E \bigl(\tilde{f}_i(\mathcal{M})),
\end{equation}
where $\varphi_E$ is an MLP and $\tilde{f}(\mathcal{M}) \in \mathbb{R}^{N \times d_2}$ denotes the atom-wise representations of $\mathcal{M}$.
Atomic forces $\hat{\mathbf{F}}_{\mathrm{force}} \in \mathbb{R}^{N \times 3}$ are then obtained as the negative gradient of the predicted energy with respect to atomic Cartesian coordinates:
\begin{equation}
    \hat{\mathbf{F}}_{\mathrm{force}} = -\nabla_{\mathbf{R}} \hat{E}, \quad \quad \mathbf{R} = \mathbf{F}\tilde{\mathbf{L}}.
\end{equation}
These models are typically trained with a combination of energy and force losses:
\begin{equation}
\mathcal{L}_{\mathrm{MLIP}} =  \lambda_{\mathrm{ener}} \mathcal{L}_{\mathrm{ener}}(\hat{E}, E) + \lambda_{\mathrm{force}} \mathcal{L}_{\mathrm{force}}(\hat{\mathbf{F}}_{\mathrm{force}}, \mathbf{F}_{\mathrm{force}}).
\end{equation}
Because accurate energy and force prediction requires sensitivity to energy-relevant structural variations, MLIP atom-wise representations provide a natural source of stability-aware supervision for guiding crystal generative models.

\section{CrystalREPA}

\begin{figure}[t]
  \centering
  \includegraphics[width=\textwidth]{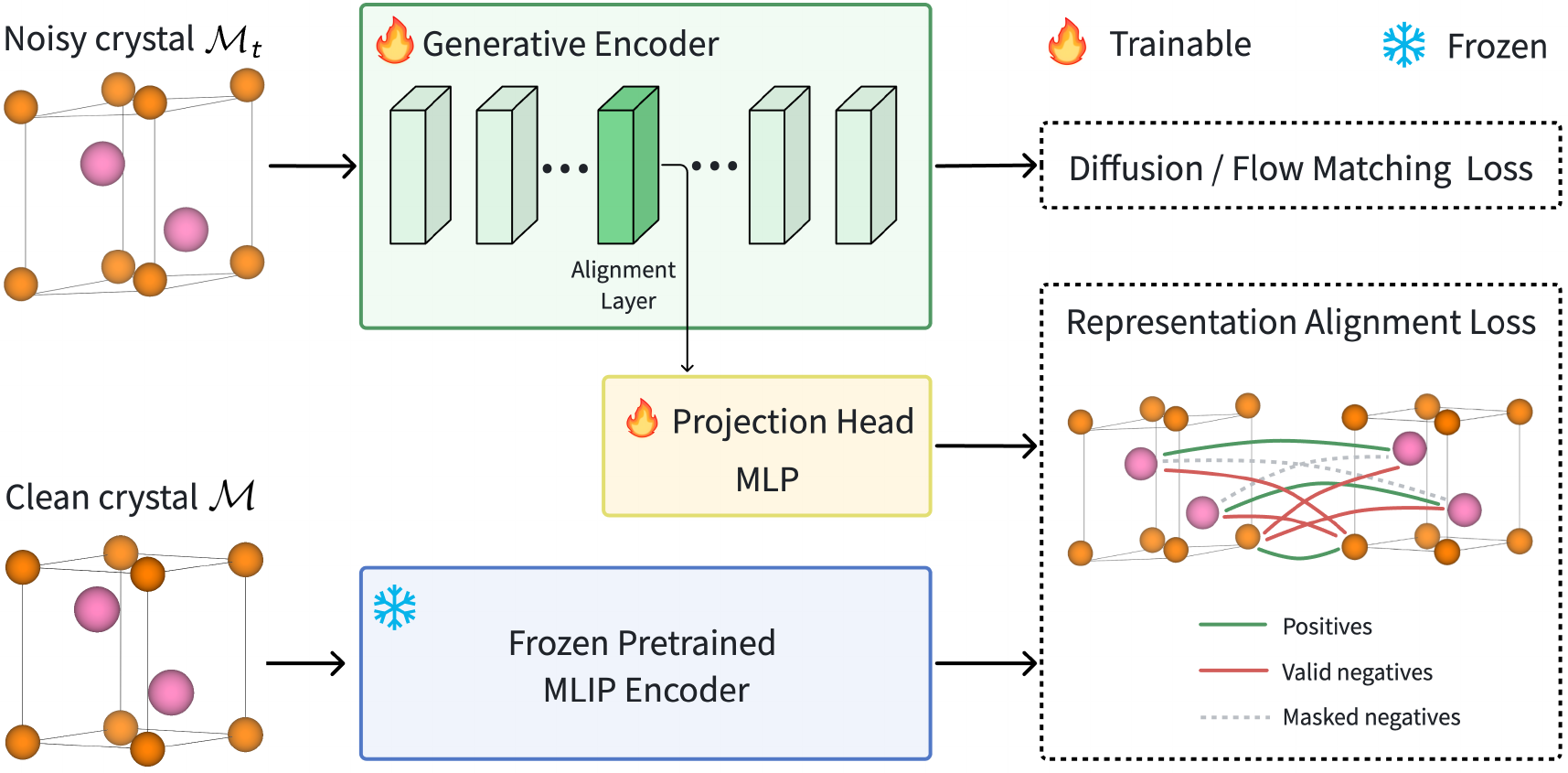}
  \caption{
  Overview of CrystalREPA. 
  During training, CrystalREPA aligns atom-wise hidden states from the generative encoder with frozen MLIP representations of the corresponding clean crystal through a projection head and an element-aware contrastive loss.
  The projection head and MLIP teacher representations are used only during training, leaving the original sampling procedure unchanged during inference.
  }
  \label{fig:method}
\end{figure}

\textbf{Motivation.}
Existing crystal generative models learn the \emph{geometry} of stable crystals, but their training objectives never expose them to the \emph{potential energy surface} that determines stability.
Motivated by the representation gap revealed by energy probing in Fig.~\ref{fig:summary}(a) and the qualitative failures in Fig.~\ref{fig:summary}(b), we propose Crystal REPresentation Alignment (CrystalREPA), a simple training framework that transfers stability-aware priors from pretrained universal MLIPs to crystal generative models.

\textbf{Core idea.}
The central idea is to align the atom-wise hidden states of a generative model on a corrupted intermediate structure $\mathcal{M}_t = (\mathbf{A}_t, \mathbf{F}_t, \mathbf{L}_t)$ with the atom-wise representations of a pretrained universal MLIP on the corresponding clean crystal $\mathcal{M} = (\mathbf{A}, \mathbf{F}, \mathbf{L})$. 
The clean crystal is used as the MLIP input to provide a stability-aware target for the generative path.
This choice is in line with the objective of learning the transport or denoising direction from $\mathcal{M}_t$ toward the clean structure.
Appendix~\ref{sec:teacher_input_ablation} empirically validates this design.
In this way, the generative encoder is encouraged to learn features that are not only useful for generation, but also consistent with representation patterns learned from large-scale energy and force supervision.

Formally, we align rotationally invariant atom-wise representations from both the generative encoder and the MLIP teacher.
Let $\tilde{\mathbf{z}} = \tilde{f}(\mathcal{M}) \in \mathbb{R}^{N \times d_2}$ denote the invariant atom-wise teacher representations extracted from a pretrained universal MLIP, which is kept frozen throughout training.
Let $\mathbf{h}_t^{(k)} = f^{(k)}(\mathcal{M}_t, t) \in \mathbb{R}^{N \times d_1}$ denote the invariant atom-wise hidden states of the generative model at layer $k$.
Since the teacher and student representations generally live in different latent spaces, we introduce a lightweight projection head $\varphi_p$ to map the student features into the teacher space: $\mathbf{z} = \varphi_p (\mathbf{h}_t^{(k)}) \in \mathbb{R}^{N \times d_2}$.
Here, $\varphi_p$ is implemented as an MLP and is only used during training.

In practice, we compute the contrastive alignment objective over all atoms in a mini-batch rather than within each crystal independently.
Let $\mathcal{I}$ denote the flattened set of atoms from all crystals in the mini-batch, and let $N_B=|\mathcal{I}|$ be the total number of atoms.
Because $\mathcal{M}_t$ is constructed by corrupting the clean crystal $\mathcal{M}$ without permuting atoms, the atom ordering is preserved within each crystal.
Thus, for each atom $i\in\mathcal{I}$, the projected student representation $\mathbf{z}_i$ and teacher representation $\tilde{\mathbf{z}}_i$ form a natural positive pair.
We compute the cosine similarity matrix $\mathbf{S}\in\mathbb{R}^{N_B\times N_B}$ between all projected student representations and teacher representations in the mini-batch:
\begin{equation}
\label{eq:s_ij}
\mathbf{S}_{ij}
=
\frac{\mathbf{z}_i^\top \tilde{\mathbf{z}}_j}
{\|\mathbf{z}_i\|\|\tilde{\mathbf{z}}_j\|},
\quad i,j\in\mathcal{I}.
\end{equation}
A standard contrastive objective would treat all off-diagonal pairs as negatives~\cite{oord2018representation}.
However, this is suboptimal for crystals: atoms of the same element often share similar chemical characteristics and may therefore be naturally close in the teacher representation space.
Treating such pairs as negatives introduces false-negative supervision and can undesirably distort the representation geometry.
To mitigate this issue, we introduce an element-aware mask $\mathbb{M}$, constructed from the clean atom types $\mathbf{A}$, that removes same-element off-diagonal pairs from the negative set while keeping diagonal positive pairs unchanged.
\begin{equation}
\label{eq:mask}
    \mathbb{M}_{ij} =
    \begin{cases}
    0, & i \neq j \text{ and } \mathbf{A}_i = \mathbf{A}_j, \\
    1, & \text{otherwise.}
    \end{cases}
\end{equation}
Based on this mask, we define the representation alignment loss as a symmetric Element-Aware InfoNCE (EA-NCE) objective in Eq.~\eqref{eq:eance}, where $\tau$ is a temperature hyperparameter.
This objective encourages atom-wise consistency between student representations and teacher representations, while avoiding unnecessary repulsion between chemically similar atoms of the same species.
\begin{equation}
\label{eq:eance}
    \mathcal{L}_{\mathrm{REPA}} = -\frac{1}{2{N_B}} \sum_{i=1}^{N_B} (\log \frac{\exp(\mathbf{S}_{ii} / \tau)}{\sum_{j=1}^{N_B} \mathbb{M}_{ij} \exp(\mathbf{S}_{ij} / \tau)} + \log \frac{\exp(\mathbf{S}_{ii} / \tau)}{\sum_{j=1}^{N_B} \mathbb{M}_{ji}\exp(\mathbf{S}_{ji} / \tau)}).
\end{equation}

Finally, we combine the alignment objective with the original generative loss. 
The resulting training objective contains both the original generation loss and the representation alignment loss, weighted by a hyperparameter $\lambda_{\mathrm{REPA}}$ that controls the alignment strength.
\begin{equation}
\label{eq:total_loss}
    \mathcal{L} = \mathcal{L}_{\mathrm{gen}} + \lambda_{\mathrm{REPA}} \mathcal{L}_{\mathrm{REPA}}.
\end{equation}

\textbf{Computational cost.}
CrystalREPA introduces marginal training overhead and no extra inference cost.
In practice, teacher representations of clean crystals are precomputed offline, so extra training overhead comes only from the lightweight projection head and the alignment loss.
Tab.~\ref{tab:cost} shows that CrystalREPA increases the training time by only 1.0\% on average.
At inference time, both the teacher encoder and the projection head are discarded, preserving the original sampling procedure.

\begin{table}
  \caption{
  Performance comparison of CrystalREPA on MP-20. We evaluate robustness across three crystal generative frameworks and 10 universal MLIP teachers.
  "Avg. over 10 MLIPs" reports the average metric value across all MLIP teachers.
  Values in parentheses denote standard errors over five independent sampling and evaluation runs.
  }
  \label{tab:result}
  \centering
  \begin{spacing}{1.2}
  \tiny
  \begin{tabular}
{L{1.1cm} L{1.73cm} C{0.9cm}  C{0.8cm} C{0.55cm} C{0.55cm} C{0.88cm} C{0.7cm} C{0.6cm} C{0.7cm} C{0.7cm}}
    \toprule
    Model & Target Rep. & Metastable\%$\uparrow$ & MSUN\%$\uparrow$ & Stable\%$\uparrow$ & SUN\%$\uparrow$ & $\bar{E}_{\mathrm{hull}}\downarrow$ & RMSD$\downarrow$ & Valid\%$\uparrow$ & Unique\%$\uparrow$ & Novel\%$\uparrow$  \\

\midrule
    \multicolumn{2}{l}{Vanilla CrystalFlow~\cite{luo2025crystalflow}} & 41.2(0.5) & 17.2(0.5) & 5.1(0.3) & 1.2(0.1) & 0.198(0.003) & 0.60(0.01) & 84.9(0.6) & 99.6(0.2) & \textbf{73.1}(0.4) \\
\rowcolor{gray!5} +CrystalREPA & Avg. over 10 MLIPs & 50.5 & 19.1 & 6.2 & 1.5 & 0.178 & 0.51 & 86.1 & 99.7 & 65.0 \\ 
    \rowcolor{gray!5} (Ours) & DPA-3.1-3M & \textbf{55.0}(0.7) & \textbf{20.0}(0.1) & \textbf{7.0}(0.4) & 1.4(0.1) & 0.179(0.005) & 0.46(0.01) & 86.6(0.4) & 99.5(0.1) & 61.3(0.7) \\ 
    \rowcolor{gray!5}  & DPA-3.1-3M-FT & 52.5(0.7) & 18.6(0.5) & \textbf{7.0}(0.5) & 1.5(0.2) & \textbf{0.174}(0.006) & 0.55(0.01) & \textbf{87.5}(0.1) & 99.6(0.1) & 62.7(0.4) \\
    \rowcolor{gray!5}  & DPA-3.1-MPtrj & 51.5(0.4) & 19.8(0.2) & 6.2(0.7) & 1.7(0.2) & \textbf{0.174}(0.004) & 0.53(0.00) & 86.7(1.0) & 99.7(0.0) & 65.0(0.5) \\ 
    \rowcolor{gray!5}  & MACE-MPA-0 & 47.8(0.5) & 19.6(0.5) & 5.8(0.2) & 1.4(0.1) & 0.176(0.005) & \textbf{0.42}(0.00) & 84.6(0.6) & \textbf{99.9}(0.1) & 68.2(0.3) \\ 
    \rowcolor{gray!5}  & MACE-MP-0 & 49.1(0.5) & 18.7(0.6) & 5.9(0.3) & 1.5(0.2) & 0.178(0.006) & 0.54(0.01) & 85.4(0.5) & 99.7(0.0) & 65.5(0.9) \\ 
    \rowcolor{gray!5}  & ORB v3 & 50.1(0.7) & 18.0(0.5) & 6.2(0.2) & 1.4(0.1) & \textbf{0.174}(0.006) & 0.57(0.01) & 86.3(0.8) & 99.7(0.1) & 63.8(0.5) \\ 
    \rowcolor{gray!5}  & ORB v2 & 51.5(0.4) & \textbf{20.0}(0.4) & 6.2(0.5) & 1.6(0.1) & 0.178(0.005) & 0.46(0.01) & 85.6(0.7) & 99.8(0.0) & 64.6(0.6)  \\
    \rowcolor{gray!5}  & SevenNet-Omni-i12 & 46.5(0.6) & 18.5(0.5) & 5.7(0.3) & 1.4(0.1) & 0.190(0.005) & 0.59(0.01) & 85.2(0.7) & 99.7(0.0) & 69.0(1.0)  \\ 
    \rowcolor{gray!5}  & SevenNet-l3i5 & 49.5(0.4) & 19.0(0.2) & 5.7(0.2) & 1.3(0.2) & 0.176(0.005) & 0.51(0.00) & 86.6(0.6) & 99.7(0.0) & 65.7(0.4)  \\
    \rowcolor{gray!5}  & MatterSim-v1-5M & 51.3(0.7) & 19.4(0.8) & 6.5(0.2) & \textbf{1.8(0.1)} & 0.179(0.006) & 0.51(0.01) & 86.2(0.8) & 99.8(0.1) & 64.2(0.6)  \\ 
    \midrule
    \addlinespace[0.5em]
    \multicolumn{2}{l}{Vanilla MatterGen~\cite{zeni2025generative}} & 57.4(0.5) & 25.3(0.2) & 6.7(0.4) & 1.9(0.2) & 0.148(0.003) & 0.12(0.00) & 84.7(0.5) & 99.6(0.0) & \textbf{64.6}(0.7)\\
\rowcolor{gray!5} +CrystalREPA & Avg. over 10 MLIPs & 62.7 & 27.7 & 7.8 & 2.5 & 0.131 & 0.11 & 85.6 & 99.5 & 61.8  \\ 
    \rowcolor{gray!5} (Ours) & DPA-3.1-3M & 64.3(0.7) & 28.6(0.7) & 7.4(0.3) & 2.3(0.2) & 0.129(0.004) & \textbf{0.11}(0.00) & 85.8(0.4) & 99.3(0.1) & 60.7(0.9)  \\ 
    \rowcolor{gray!5}  & DPA-3.1-3M-FT & 63.7(0.7) & 27.6(0.4) & \textbf{8.7}(0.4) & 2.7(0.1) & 0.125(0.003) & \textbf{0.11}(0.00) & 85.5(0.6) & 99.5(0.1) & 60.5(0.6) \\
    \rowcolor{gray!5}  & DPA-3.1-MPtrj & 62.5(0.3) & 28.0(0.8) & 7.3(0.2) & 2.4(0.2) & 0.127(0.003) & 0.12(0.00) & 84.5(0.2) & 99.5(0.1) & 62.7(0.8)\\
    \rowcolor{gray!5}  & MACE-MPA-0 & 61.3(0.6) & 28.4(0.7) & 7.8(0.1) & 2.6(0.1) & 0.138(0.003) & \textbf{0.11}(0.00) & 85.2(0.6) & \textbf{99.7}(0.1) & 63.5(0.5) \\
    \rowcolor{gray!5}  & MACE-MP-0 & 61.3(0.5) & 27.6(0.5) & 7.3(0.2) & 2.4(0.2) & 0.135(0.004) & \textbf{0.11}(0.00) & 85.1(0.5) & 99.5(0.1) & 63.1(0.5)  \\
    \rowcolor{gray!5}  & ORB v3 & \textbf{66.2}(0.4) & \textbf{29.2}(0.4) & 8.2(0.3) & 2.6(0.2) & \textbf{0.119}(0.004) & \textbf{0.11}(0.00) & 86.5(0.6) & 99.5(0.1) & 59.9(0.3)  \\
    \rowcolor{gray!5}  & ORB v2 & 60.9(0.7) & 26.8(0.3) & 7.7(0.6) & 2.5(0.2) & 0.136(0.006) & 0.14(0.00) & 86.1(0.2) & 99.5(0.1) & 63.2(0.5)  \\
    \rowcolor{gray!5}  & SevenNet-Omni-i12 & 59.5(0.5) & 25.7(0.8) & 7.3(0.3) & 2.0(0.1) & 0.143(0.007) & \textbf{0.11}(0.00) & 85.7(0.5) & 99.4(0.1) & 62.8(0.7) \\
    \rowcolor{gray!5}  & SevenNet-l3i5 & 63.1(0.7) & 27.0(0.5) & 7.9(0.6) & 2.5(0.2) & 0.133(0.004) & \textbf{0.11}(0.00) & \textbf{86.6}(0.3) & 99.6(0.1) & 60.3(0.4)  \\
    \rowcolor{gray!5}  & MatterSim-v1-5M & 64.1(0.5) & 28.5(0.6) & 8.4(0.3) & \textbf{2.9}(0.2) & 0.127(0.001) & \textbf{0.11}(0.00) & 85.4(0.6) & 99.6(0.0) & 61.5(0.5)  \\
    \midrule
\addlinespace[0.5em]
    \multicolumn{2}{l}{Vanilla DiffCSP~\cite{jiao2023crystal}}  & 43.2(0.7) & 18.8(0.5) & 5.9(0.3) & 2.0(0.2) & 0.208(0.005) & 0.36(0.01) & 82.7(0.6) & 98.9(0.1) & \textbf{72.0}(0.4)\\
\rowcolor{gray!5} +CrystalREPA & Avg. over 10 MLIPs & 50.9 & 23.4 & 7.2 & 2.6 & 0.166 & 0.37 & 84.3 & 99.2 & 69.3 \\
    \rowcolor{gray!5} (Ours) & DPA-3.1-3M & 50.6(0.8) & 23.9(0.8) & 7.5(0.7) & 2.8(0.3) & 0.166(0.006) & 0.37(0.01) & 84.1(0.4) & 99.2(0.1) & 70.0(1.0) \\
    \rowcolor{gray!5}  & DPA-3.1-3M-FT & 52.0(0.8) & 24.3(0.7) & \textbf{8.2}(0.2) & \textbf{2.9}(0.3) & 0.165(0.006) & 0.36(0.00) & 84.5(0.6) & 99.2(0.1) & 68.6(0.6) \\
    \rowcolor{gray!5}  & DPA-3.1-MPtrj & 50.2(0.4) & 24.7(0.7) & 6.7(0.1) & 2.5(0.1) & 0.175(0.004) & 0.40(0.01) & 84.9(0.5) & \textbf{99.5}(0.1) & 71.4(0.6)  \\
    \rowcolor{gray!5}  & MACE-MPA-0 & 49.9(0.8) & 23.4(0.5) & 6.0(0.4) & 2.1(0.2) & 0.169(0.004) & 0.38(0.01) & 84.8(0.9) & 99.2(0.1) & 70.4(0.6)  \\
    \rowcolor{gray!5}  & MACE-MP-0 & 49.8(0.6) & 23.4(0.4) & 6.4(0.5) & 2.5(0.3) & 0.166(0.004) & 0.37(0.01) & 84.4(0.8) & 99.2(0.1) & 70.4(0.5)  \\
    \rowcolor{gray!5}  & ORB v3 & \textbf{56.5}(0.8) & \textbf{24.8}(0.8) & 8.1(0.3) & 2.5(0.1) & \textbf{0.147}(0.003) & \textbf{0.32}(0.00) & 84.5(0.7) & 99.2(0.1) & 65.3(0.4)  \\
    \rowcolor{gray!5}  & ORB v2 & 53.5(1.0) & 23.2(1.1) & 8.1(0.1) & 2.7(0.1) & 0.158(0.007) & 0.36(0.01) & 84.9(0.6) & 99.2(0.0) & 66.2(0.7)  \\
    \rowcolor{gray!5}  & SevenNet-Omni-i12 & 45.8(0.7) & 21.0(0.8) & 6.6(0.5) & 2.6(0.1) & 0.180(0.005) & 0.35(0.01) & 83.2(0.4) & 98.8(0.1) & \textbf{72.0}(0.5)  \\
    \rowcolor{gray!5}  & SevenNet-l3i5 & 47.7(0.6) & 22.0(0.7) & 6.3(0.4) & 2.6(0.2) & 0.180(0.005) & 0.40(0.00) & 82.7(0.6) & 99.3(0.2) & 70.6(0.3)  \\
    \rowcolor{gray!5}  & MatterSim-v1-5M & 52.6(0.5) & 23.7(0.3) & 7.9(0.3) & \textbf{2.9}(0.2) & 0.158(0.004) & 0.36(0.01) & \textbf{85.4}(0.7) & 99.4(0.2) & 68.0(0.7) \\
    
    \bottomrule
  \end{tabular}
  \end{spacing}
\end{table}

\section{Experiments}
\label{sec:experiments}

We evaluate CrystalREPA by addressing the following research questions:
(1) Does CrystalREPA narrow the representation gap between generative models and universal MLIPs? (Fig.~\ref{fig:summary}a);
(2) Does CrystalREPA consistently improve crystal generation quality across different generative frameworks, teacher MLIPs, and benchmark datasets? (Fig.~\ref{fig:summary}, Tabs.~\ref{tab:result} and~\ref{tab:result_alex_mp});
(3) Does benchmark accuracy (e.g., Matbench Discovery) predict an MLIP's value as a teacher, and if not, what property does? (Fig.~\ref{fig:rep_norm}); and
(4) How do individual components of CrystalREPA contribute to its overall performance? (Fig.~\ref{fig:ablation}).

\subsection{CrystalREPA narrows the representation gap}
To evaluate how predictive generative representations are of formation energy, we perform a linear probing task on their learned representations.
Specifically, we extract the atom-wise hidden states from the encoders of various models using clean structures ($t=1$) from the MP-20 dataset. 
These hidden states are aggregated into global structural representations via mean pooling, which then serve as inputs for a ridge regression model to predict the formation energy.
As an MLIP reference, we apply the same probing protocol to DPA-3.1-3M.
See Appendix~\ref{sec:formation_energy_probing} for details.

Fig.~\ref{fig:summary}(a) shows that the representations learned by vanilla CrystalFlow exhibit a significant representation gap compared to the pretrained MLIP DPA-3.1-3M, characterized by substantially higher Mean Absolute Errors (MAE) across all encoder layers.
CrystalREPA significantly narrows this gap, with the probing error in the intermediate layers approaching that of the universal MLIP, suggesting that representation alignment makes the generative representations more predictive of stability-related quantities.
This improvement is also reflected qualitatively in Fig.~\ref{fig:summary}(b). 
In the examples shown, starting from the same noisy structures, vanilla CrystalFlow can generate structurally or compositionally invalid crystals, whereas CrystalREPA produces novel and metastable structures.
Together with broader probing results in Appendix~\ref{sec:additional_probe}, these results suggest that CrystalREPA narrows the representation gap and that the improved representation
quality is associated with more stable and valid generated structures.
Across three generative frameworks and their CrystalREPA variants, formation-energy probing MAE is also negatively correlated with downstream Metastable\% (Pearson $r=-0.62$), as detailed in Appendix~\ref{app:additional_representation}.

\subsection{CrystalREPA consistently improves crystal generation quality}

\begin{table}
  \caption{
  Performance comparison of CrystalREPA on Alex-MP-20.
  We use DPA-3.1-3M as the MLIP teacher and evaluate CrystalREPA across three crystal generative frameworks.
  Values in parentheses denote standard errors over five independent sampling and evaluation runs.
  }
  \label{tab:result_alex_mp}
  \centering
  \begin{spacing}{1.2}
  \tiny
  \begin{tabular}
    {L{1.1cm} L{1.3cm} C{0.9cm}  C{0.8cm} C{0.6cm} C{0.6cm} C{0.9cm} C{0.7cm} C{0.7cm} C{0.7cm} C{0.7cm}}
    \toprule
    Model & Target Rep. & Metastable\%$\uparrow$ & MSUN\%$\uparrow$ & Stable\%$\uparrow$ & SUN\%$\uparrow$ & $\bar{E}_{\mathrm{hull}}\downarrow$ & RMSD$\downarrow$ & Valid\%$\uparrow$ & Unique\%$\uparrow$ & Novel\%$\uparrow$  \\
    \midrule
\multicolumn{2}{l}{Vanilla CrystalFlow~\cite{luo2025crystalflow}} & 61.8(0.8) & 31.5(0.6) & \textbf{4.8}(0.1) & 1.6(0.3) & 0.100(0.002) & 0.36(0.00) & 87.7(0.3) & \textbf{100.0}(0.0) & 66.3(0.6) \\ 
\rowcolor{gray!5} +CrystalREPA & DPA-3.1-3M & \textbf{63.0}(0.2) & \textbf{33.7}(0.6) & \textbf{4.8}(0.1) & \textbf{1.9}(0.1) & \textbf{0.097}(0.001) & \textbf{0.31}(0.01) & \textbf{88.1}(0.4) & 99.9(0.0) & \textbf{67.9}(0.6)  \\
    \midrule
    \multicolumn{2}{l}{Vanilla MatterGen~\cite{zeni2025generative}} & 72.4(0.2) & 38.1(0.7) & 6.9(0.1) & 2.9(0.3) & 0.079(0.001) & \textbf{0.07}(0.00) & \textbf{85.8}(0.8) & \textbf{100.0}(0.0) & 62.3(0.6) \\ 
    \rowcolor{gray!5} +CrystalREPA & DPA-3.1-3M & \textbf{73.0}(0.7) & \textbf{39.3}(0.5) & \textbf{7.2}(0.1) & \textbf{3.6}(0.1) & \textbf{0.075}(0.000) & \textbf{0.07}(0.00) & \textbf{85.8}(0.5) & 99.9(0.0) & \textbf{63.1}(0.5)  \\ 
    \midrule
\multicolumn{2}{l}{Vanilla DiffCSP~\cite{jiao2023crystal}} & 63.1(1.0) & 35.0(0.5) & 5.4(0.5) & 1.7(0.2) & 0.100(0.002) & \textbf{0.25}(0.01) & 87.6(0.3) & \textbf{99.8}(0.1) & 69.5(0.7) \\
    \rowcolor{gray!5} +CrystalREPA & DPA-3.1-3M & \textbf{63.4}(0.5) & \textbf{36.3}(0.6) & \textbf{6.0}(0.3) & \textbf{2.2}(0.2) & \textbf{0.097}(0.002) & 0.26(0.01) & \textbf{87.9}(0.3) & 99.6(0.1) & \textbf{70.7}(0.5)   \\ 
    \bottomrule
  \end{tabular}
  \end{spacing}
\end{table}

We next evaluate whether CrystalREPA improves crystal generation quality.
We evaluate CrystalREPA on MP-20 and Alex-MP-20, using the MatterGen evaluation pipeline~\cite{zeni2025generative}. 
For each method, we sample 1,024 structures, relax them with MatterSim-v1-1M~\cite{yang2024mattersim}, and report thermodynamic stability, structural fidelity, validity, and diversity metrics.
Metastable\% and Stable\% denote the fractions of generated structures with $E_{\mathrm{hull}} \leq 0.1$ eV/atom and $E_{\mathrm{hull}} \leq 0$ eV/atom, respectively;
MSUN\% and SUN\% further require the samples to be unique and novel.
RMSD measures structural fidelity, with lower values indicating smaller structural distortion after relaxation.
Detailed metric definitions are provided in Appendix~\ref{sec:metric_definition}.

Tab.~\ref{tab:result} shows that CrystalREPA consistently improves generation quality on MP-20 across CrystalFlow, MatterGen, and DiffCSP.
Averaged over 10 MLIP teachers, CrystalREPA improves Metastable\% from 41.2 to 50.5 for CrystalFlow, from 57.4 to 62.7 for MatterGen, and from 43.2 to 50.9 for DiffCSP. 
MSUN\% also increases from 17.2 to 19.1, 25.3 to 27.7, and 18.8 to 23.4, respectively.
These gains are accompanied by higher Stable\% and SUN\%, lower average energy above hull ($\bar{E}_{\mathrm{hull}}$), and generally improved validity, indicating that the gains are not limited to a single stability metric.
The individual teacher rows further show that the gains are robust across MLIP families, including DPA, MACE, ORB, SevenNet, and MatterSim.
The strongest MP-20 result is achieved by MatterGen + CrystalREPA with ORB v3, reaching 66.2 Metastable\%, 29.2 MSUN\%, and the lowest 
average energy above hull of 0.119 eV/atom.

Tab.~\ref{tab:result_alex_mp} shows the same trend on the larger Alex-MP-20 dataset.
The gains are smaller than on MP-20, which is expected because the Alex-MP-20 baselines are already stronger, but the direction of improvement remains consistent across all three generators.
With DPA-3.1-3M as the MLIP teacher, CrystalREPA improves MSUN\% from 31.5 to 33.7 for CrystalFlow, from 38.1 to 39.3 for MatterGen, and from 35.0 to 36.3 for DiffCSP.
These results indicate that CrystalREPA remains beneficial across datasets and base generators.

Finally, Unique\% remains nearly saturated around 99\% across most settings.
Although raw Novel\% decreases in some MP-20 cases, both MSUN\% and SUN\%, which jointly require novelty, uniqueness, and stability, still improve.
This suggests that CrystalREPA improves the yield of novel, unique, and stable structures despite a modest decrease in raw novelty in some settings.

We further assess generated crystals through DFT-based evaluation,
finite-temperature molecular dynamics simulations,
predicted synthesizability, and validity-aware MSUN.
Across these complementary evaluations, CrystalREPA
consistently improves the key metrics.
Notably, on CrystalFlow, CrystalREPA improves DFT-evaluated
Metastable\% from 50.0 to 64.0 and MSUN\% from 15.0 to 22.5.
Further details are provided in Appendix~\ref{app:additional_validation}.

\begin{figure}[t]
  \centering
  \includegraphics[width=\textwidth]{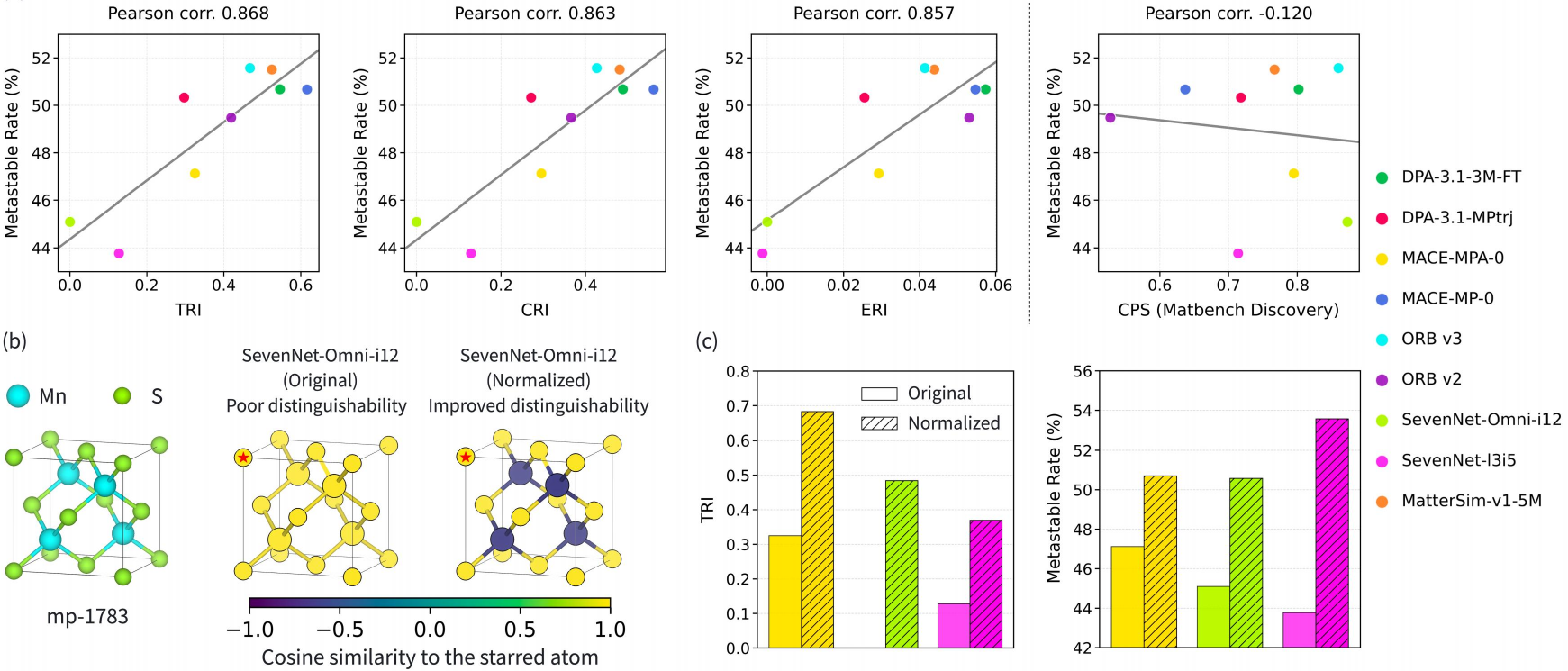}
  \caption{
Representation distinguishability is associated with differences in MLIP transfer effectiveness.
  (a) Metastable rate correlates strongly with Chemical Resolution Index (CRI), Environment Resolution Index (ERI), and Total Resolution Index (TRI), but shows little association with the Matbench Discovery Combined Performance Score (CPS).
  (b) Similarity visualization shows improved atom-wise distinguishability after representation normalization.
(c) Representation normalization increases distinguishability and is accompanied by consistent improvements in metastable rates.
  The three teacher MLIPs from left to right are MACE-MPA-0, SevenNet-Omni-i12, and SevenNet-l3i5.
}
  \label{fig:rep_norm}
\end{figure}

\subsection{Distinguishability, not benchmark accuracy, predicts transfer effectiveness}

While CrystalREPA improves over the vanilla baseline across all teachers, the magnitude of improvement varies considerably.
For example, SevenNet-Omni-i12 achieves the highest Combined Performance Score (CPS) on Matbench Discovery~\cite{riebesell2025framework} among the MLIPs considered, but yields relatively small gains in generation metrics in Tab.~\ref{tab:result}.
This mismatch suggests that benchmark performance alone does not predict an MLIP's transfer effectiveness, and the answer instead lies in the structure of its atom-wise representation space.

We therefore examine transfer effectiveness from a representation perspective.
Inspired by recent work showing that pairwise similarity
between patch tokens among teacher representations affects transfer effectiveness in image generation~\cite{singh2026what}, we quantify the distinguishability of atom-wise teacher representations using three metrics: Chemical Resolution Index (CRI), Environment Resolution Index (ERI), and Total Resolution Index (TRI). 
Intuitively, CRI measures distinguishability across elemental species, ERI measures distinguishability across local geometric environments, and TRI summarizes both aspects.
Higher values indicate a more distinguishable atom-wise representation space. 
Precise definitions are provided in Appendix~\ref{sec:cri}.
To avoid confounding teacher distinguishability with the element-aware negative sampling in EA-NCE, we conduct this diagnostic analysis using a cosine alignment objective rather than EA-NCE.

Fig.~\ref{fig:rep_norm}(a) shows that representation distinguishability strongly correlates with generative performance across the unnormalized representations of the nine MLIP teachers with available Matbench Discovery CPS scores.
The metastable rate has high positive Pearson correlations with CRI, ERI, and TRI, reaching 0.863, 0.857, and 0.868, respectively. 
In contrast, CPS from Matbench Discovery shows little positive association, with a correlation of -0.120.
This helps explain why a teacher MLIP can perform well on conventional predictive benchmarks yet provide limited benefit for CrystalREPA if its atom-wise representation space does not clearly resolve chemical identity and local environment.
We further introduce the Energy Resolution Index (EnRI) to quantify the distinguishability of MLIP representations across thermodynamic stability levels.
EnRI also exhibits a strong correlation with downstream Metastable\% (Pearson $r=0.936$), as detailed in Appendix~\ref{app:additional_representation}.

Fig.~\ref{fig:rep_norm}(b, c) further suggests that improved teacher representation distinguishability is associated with better transfer effectiveness.
We apply a simple representation normalization, defined in Appendix~\ref{sec:rep_norm}, to frozen teacher representations of MACE-MPA-0, SevenNet-Omni-i12 and SevenNet-l3i5 before alignment.
As shown in Fig.~\ref{fig:rep_norm}(b), for SevenNet-Omni-i12, normalization induces sharper contrast between sulfur atoms and manganese atoms, illustrating that the transformed representation space makes chemically distinct atoms more distinguishable.
Quantitatively, as illustrated in Fig.~\ref{fig:rep_norm}(c), 
normalization increases TRI for all three teachers with relatively low original distinguishability and is accompanied by consistent improvements in metastable rate. 
For example, when using SevenNet-Omni-i12 as the MLIP teacher, TRI increases from 0.00 to 0.48, and the metastable rate improves from 45.1 to 50.6.
Beyond increasing distinguishability, normalization also changes the scale and geometry of the alignment targets and may affect optimization, so improvements in distinguishability may not translate proportionally into downstream performance gains.

Overall, Fig.~\ref{fig:rep_norm} suggests that, in our experiments, the transfer effectiveness of a teacher MLIP is better predicted by the distinguishability of its atom-wise representations than by benchmark performance,
providing a practical criterion for selecting or preprocessing future teachers for crystal generation.

\subsection{Ablation studies}
\begin{figure}[t]
  \centering
  \includegraphics[width=\textwidth]{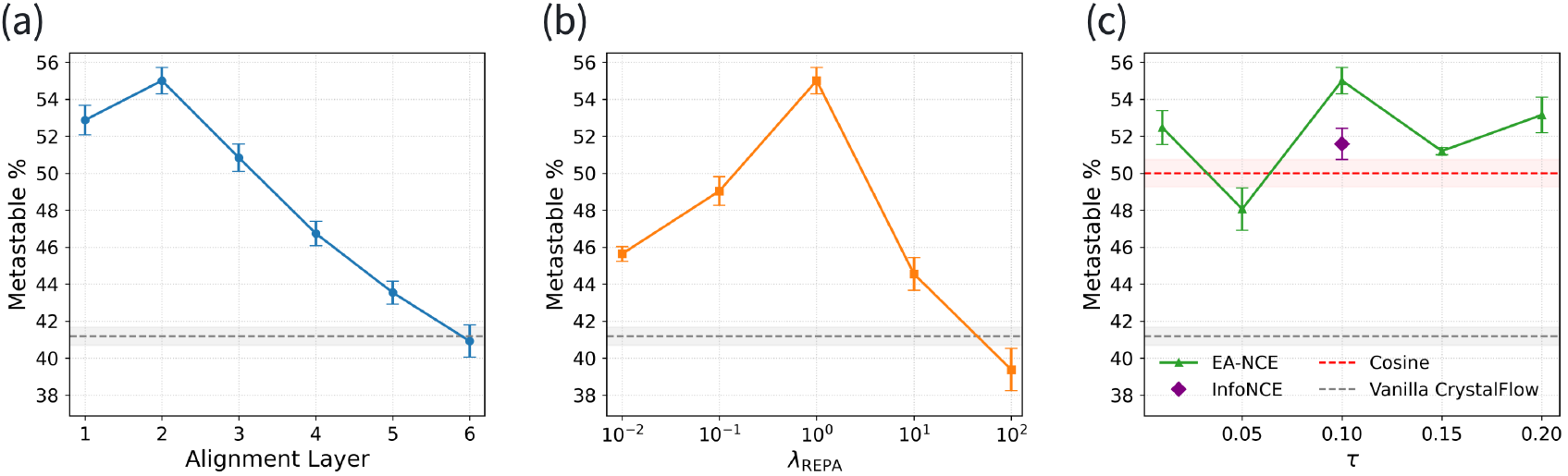}
  \caption{
  Ablation studies of CrystalREPA using CrystalFlow as the base generative model and DPA-3.1-3M as the MLIP teacher. 
  We study the effects of (a) the alignment layer in the generative model, (b) the alignment weight $\lambda_{\mathrm{REPA}}$, and (c) the choice of alignment objective.
  }
  \label{fig:ablation}
\end{figure}

We further conduct ablation studies to validate the design of CrystalREPA and examine its robustness to hyperparameter choices.
All results in Fig.~\ref{fig:ablation} use CrystalFlow as the base generative model and DPA-3.1-3M as the MLIP teacher. 
Overall, CrystalREPA improves over the vanilla model across a broad range of configurations, indicating that its gains do not rely on delicate hyperparameter tuning.

\textit{Alignment Layer.}
As shown in Fig.~\ref{fig:ablation}(a), aligning most encoder layers of CrystalFlow improves the metastable rate over the vanilla baseline, with the best performance achieved when aligning the hidden states after the second message-passing layer.
Performance gradually declines for deeper alignment layers, suggesting that early intermediate representations provide the most effective transfer.

\textit{Effect of $\lambda_{\mathrm{REPA}}$}.
Fig.~\ref{fig:ablation}(b) shows that CrystalREPA remains effective for $\lambda_{\mathrm{REPA}}$ from 0.01 to 10, achieves the best performance at $\lambda_{\mathrm{REPA}} = 1$, and degrades when the alignment weight becomes excessively large.
This indicates that CrystalREPA is robust to moderate choices of the loss weight.

\textit{Alignment objective.}
Fig.~\ref{fig:ablation}(c) compares different alignment objectives.
The proposed EA-NCE consistently outperforms vanilla CrystalFlow across different temperatures ($\tau$) from 0.01 to 0.2, with the best result at $\tau = 0.1$.
It also performs better than standard InfoNCE~\cite{oord2018representation} and the negative cosine similarity objective, validating the element-aware masking design for avoiding false negatives between same-element atoms.

\textit{Compared to conditional generation.}
Tab.~\ref{tab:result_condition} shows that energy-conditioned generation improves some stability-oriented metrics, such as $\bar{E}_{\mathrm{hull}}$, Stable\%, and SUN\%, but consistently reduces Valid\% and therefore harms the physical plausibility of generated structures.
In contrast, CrystalREPA yields more balanced improvements in Metastable\%, MSUN\%, RMSD, and Valid\%, leading to stronger overall generation quality.
Detailed results are deferred to Appendix~\ref{sec:condition}.

\textit{Effect of data overlap.}
We also analyze the potential effect of overlap between the MLIP pretraining data and the generator training data.
Appendix~\ref{sec:dataoverlap} shows that CrystalREPA remains beneficial with a non-overlap teacher and achieves comparable gains with a high-overlap teacher, suggesting that dataset overlap alone is unlikely to be the primary driver of the observed improvements.

\textit{Additional ablations.}
We further investigate SOAP-based alignment, energy-direction removal, MLIP teacher-layer selection, and direct MLIP-only optimization.
These additional experiments provide complementary evidence for the contribution of energy-related MLIP representations and the importance of structural priors learned by generative models.
Detailed results are provided in Appendix~\ref{app:additional_ablations}.

\section{Conclusion}
\label{sec:conclusion}
In this work, we identified a clear representation gap between crystal generative models and pretrained universal MLIPs, and proposed CrystalREPA, a framework that transfers stability-aware priors to crystal generative models via atom-wise representation alignment.
Across three generative frameworks, ten MLIP teachers, and two benchmark datasets, CrystalREPA consistently improves thermodynamic stability, structural validity, and structural fidelity, while preserving strong diversity, introducing only marginal training overhead, and incurring no additional inference cost.
Beyond these empirical gains, we find that transfer effectiveness is predicted by the distinguishability of teacher atomistic representations rather than by benchmark accuracy, yielding a practical criterion for selecting MLIP teachers.
We hope this work provides a useful step toward bridging generative modeling and atomistic representation learning, and motivates future generative models that produce structures of higher thermodynamic stability for materials discovery.

\textbf{Limitation.}
First, the current framework aligns only atom-level representations and does not consider edge-level or higher-order structural representations. 
Second, our experiments are limited to crystalline materials, and we do not test whether the same alignment strategy transfers to other atomistic domains such as molecules or proteins. 
We leave these directions to future work.

\begin{ack}
We acknowledge the AI for Science Institute for providing the computational resources used in this work.
The work of H.W.~is supported by the National Natural Science Foundation of China (Grants No.~12525113 and No.~12561160120).
Tiejun Li acknowledges the support from National Key R\&D Program of China under grant 2021YFA1003301, and National Science Foundation of China under grant 12288101.

\noindent\textbf{Competing interests.}
The authors declare no competing interests.
\end{ack}

\bibliographystyle{unsrtnat}
\bibliography{ref}

\newpage
\appendix

\clearpage
\enlargethispage{2\baselineskip}
\tableofcontents
\clearpage

\newpage

\section{Formulation of CrystalREPA}
\subsection{CrystalREPA in flow matching frameworks}
\label{sec:crystalflow}
\textbf{Flow matching training.}
We use CrystalFlow as a representative flow-based crystal generative framework.
We denote a crystal with $N$ atoms in the unit cell as $\mathcal{M} = (\mathbf{A}, \mathbf{F}, \mathbf{L})$, where $\mathbf{A} = [\mathbf{a}_1^\top; \mathbf{a}_2^\top; \dots; \mathbf{a}_N^\top] \in \mathbb{R}^{N \times a}$ denotes atom types, with each atom type mapped to an $a$-dimensional categorical representation.
$\mathbf{F} =[\mathbf{f}_1^\top; \mathbf{f}_2^\top; \dots; \mathbf{f}_N^\top] \in [0,1)^{N \times 3}$ denotes fractional coordinates, and $\mathbf{L} \in \mathbb{R}^{6}$ denotes a rotation-invariant lattice parameterization derived from the lattice matrix $\tilde{\mathbf{L}} = [\mathbf{l}_1^\top; \mathbf{l}_2^\top;\mathbf{l}_3^\top] \in \mathbb{R}^{3 \times 3}$ via polar decomposition as $\tilde{\mathbf{L}}^\top = \mathbf{Q}\mathrm{exp}(\sum_{i=1}^6 \mathbf{L}_i \mathbf{B}_i)$.
Here, $\mathbf{Q}$ is an orthogonal matrix representing rotational degrees of freedom, $\mathrm{exp}(\cdot)$ denotes the matrix exponential, and $\{\mathbf{B}_i\}_{i=1}^6 \in \mathbb{R}^{3 \times 3}$ forms a standard basis of symmetric matrices.
In CrystalFlow, crystal generation is formulated as learning a continuous transport map from a simple prior distribution to the data distribution.

Specifically, let $\mathcal{M}_1 = (\mathbf{A}_1, \mathbf{F}_1, \mathbf{L}_1)$ be a crystal sampled from the data distribution and let $\mathcal{M}_0 = (\mathbf{A}_0, \mathbf{F}_0, \mathbf{L}_0)$ be sampled from the corresponding priors.
A continuous path $\mathcal{M}_t = (\mathbf{A}_t, \mathbf{F}_t, \mathbf{L}_t)$, with $t \in [0,1]$, interpolates between the prior and data, where $t=0$ corresponds to the prior and $t = 1$ corresponds to the clean crystal.
The interpolations for atom types and lattice variables are defined by linear interpolation between the prior sample and the data sample:
\begin{equation}
\label{eq:inter_A}
    \mathbf{A}_t = t\mathbf{A}_1 + (1-t)\mathbf{A}_0, \quad \mathbf{A}_0 \sim \mathcal{N}(0, \mathbf{I}) \in \mathbb{R}^{N \times a},
\end{equation}
\begin{equation}
\label{eq:inter_L}
  \mathbf{L}_t=t\mathbf{L}_1+(1-t)\mathbf{L}_0,
  \quad
  \mathbf{L}_0 \sim 
  \mathcal{N}\left(
  (0,0,0,0,0,1),
  0.1^2\mathbf{I}
  \right)\in \mathbb{R}^6.
\end{equation}
Fractional coordinates lie on a three-dimensional torus and must respect periodic translation invariance.
CrystalFlow addresses this by using a uniform prior and determining $\mathbf{F}_t$ with the minimum image convention:
\begin{equation}
\label{eq:inter_F}
    \mathbf{F}_t = \mathbf{F}_0 + t(w(\mathbf{F}_1 - \mathbf{F}_0 - 0.5) - 0.5), \quad \mathbf{F}_0 \sim \mathcal{U}([0,1)^{N\times 3}),
\end{equation}
where $w(\mathbf{x}) = \mathbf{x} - \lfloor \mathbf{x} \rfloor$ is the wrapping operation.

Given this probability path, the target velocity fields for atom types, fractional coordinates, and lattice variables are
\begin{equation}
\label{eq:tar_v}
    \mathbf{u}_t^A = \mathbf{A}_1 - \mathbf{A}_0, \quad \quad
    \mathbf{u}_t^F = w(\mathbf{F}_1 - \mathbf{F}_0 - 0.5)-0.5, \quad \quad
    \mathbf{u}_t^L = \mathbf{L}_1 - \mathbf{L}_0.
\end{equation}
CrystalFlow learns three time-dependent vector fields, $\mathbf{v}_\theta^A(\mathcal{M}_t, t)$,  $\mathbf{v}_\theta^F(\mathcal{M}_t, t)$, and $\mathbf{v}_\theta^L(\mathcal{M}_t, t)$, to regress these targets.
These vector fields are parameterized by a periodic E(3)-equivariant graph neural network with $K$ message-passing layers.
Let $f^{(k)}(\mathcal{M}_t, t) \in \mathbb{R}^{N \times d_1}$ denote atom-wise hidden states after the $k$-th message-passing layer.
The original CrystalFlow training objective is the weighted regression loss defined in Eqs.~\eqref{eq:predict}--\eqref{eq:loss_gen}.

\textbf{Applying CrystalREPA to flow matching models.}
CrystalREPA is applied to CrystalFlow by aligning the atom-wise hidden states along the flow path with MLIP representations of the corresponding clean crystal.
For a corrupted structure $\mathcal{M}_t$, we extract the student representation $\mathbf{h}^{(k)}_t=f^{(k)}(\mathcal{M}_t,t)\in\mathbb{R}^{N\times d_1}$ from the $k$-th layer of the generative encoder.
The clean crystal $\mathcal{M}_1$ is fed into a frozen pretrained MLIP encoder $\tilde{f}$ to obtain the teacher representation $\tilde{\mathbf{z}}=\tilde{f}(\mathcal{M}_1)\in\mathbb{R}^{N\times d_2}$.
A lightweight projection head $\varphi_p$ maps the student representation to the teacher representation space: $\mathbf{z}=\varphi_p(\mathbf{h}^{(k)}_t)\in\mathbb{R}^{N\times d_2}$.
We therefore apply the element-aware contrastive loss defined in Eqs.~\eqref{eq:s_ij}--\eqref{eq:eance}.
The final training objective for CrystalREPA in the flow matching framework is in Eq.~\eqref{eq:total_loss}.

\textbf{Inference.}
At inference time, CrystalREPA does not change the original CrystalFlow sampling procedure.
A random initial state $\mathcal{M}_0 = (\mathbf{A}_0, \mathbf{F}_0, \mathbf{L}_0)$ is sampled from the prior distributions defined above and then evolved from $t=0$ to $t=1$ using the learned vector fields.
We discretize the interval $[0,1]$ into 100 steps with $\Delta t = 1/100$ and $t_m=m\Delta t$. 
Using the Euler solver, the updates are given by
\begin{equation}
\label{eq:Euler_A}
  \mathbf{A}_{t_{m+1}} = \mathbf{A}_{t_m} + \mathbf{v}^{A}_{\theta}(\mathcal{M}_{t_m},t_m)\Delta t,
\end{equation}
\begin{equation}
\label{eq:Euler_F}
  \mathbf{F}_{t_{m+1}} = \mathbf{F}_{t_m} + (1 + 5t_m)\mathbf{v}^{F}_{\theta}(\mathcal{M}_{t_m},t_m)\Delta t,
\end{equation}
\begin{equation}
\label{eq:Euler_L}
  \mathbf{L}_{t_{m+1}} = \mathbf{L}_{t_m} + \mathbf{v}^{L}_{\theta}(\mathcal{M}_{t_m},t_m)\Delta t,
\end{equation}
for $m=0,\ldots,99$, where $(1 + 5t_m)$ is the anti-annealing scaling factor applied to fractional coordinate updates in CrystalFlow.
The MLIP teacher and projection head are discarded during inference, so CrystalREPA introduces no additional inference cost.

\subsection{CrystalREPA in diffusion frameworks}
\label{sec:crystalrepa_diffusion}
\textbf{Diffusion training.}
We first formulate the forward corruption process and denoising objective for diffusion-based crystal generative frameworks, including MatterGen and DiffCSP.
In this section, we provide a unified formulation for applying CrystalREPA to such models, while noting that the exact diffusion parameterization of each variable is model-specific.
Different from the flow matching convention used in CrystalFlow, where $t=0$ corresponds to the prior and $t=1$ corresponds to the clean crystal, diffusion models usually define $t=0$ as the clean structure and $t=T$ as pure noise.

The forward diffusion of the continuous lattice variable $\mathbf{L}$ follows Denoising Diffusion Probabilistic Model (DDPM)~\cite{ho2020denoising}.
Given the variance schedule $\beta_1,\ldots,\beta_T$, the forward process can be expressed as:
\begin{equation}
\label{eq:diffusion_L}
  q(\mathbf{L}_t \mid \mathbf{L}_0)
=
\mathcal{N}\!\left(
\mathbf{L}_t \mid \sqrt{\bar{\alpha}_t}\mathbf{L}_0,
(1-\bar{\alpha}_t)\mathbf{I}
\right),
\quad
\mathbf{L}_t
=
\sqrt{\bar{\alpha}_t}\mathbf{L}_0
+
\sqrt{1-\bar{\alpha}_t}\boldsymbol{\epsilon}_{L},
\quad
\boldsymbol{\epsilon}_L \sim\mathcal{N}(0,\mathbf{I}),
\end{equation}
where $\alpha_t = 1 - \beta_t$ and $\bar{\alpha}_t = \prod_{s=1}^{t} \alpha_s$.

Fractional coordinates $\mathbf{F}$ are diffused using score matching~\cite{song2021scorebased} with a wrapped normal distribution~\cite{bortoli2022riemannian}.
The forward process uses a wrapped normal distribution to maintain periodic invariance:
\begin{equation}
\label{eq:diffusion_F}
  \begin{aligned}
  q(\mathbf{F}_t \mid \mathbf{F}_0)
  &=
  \mathcal{N}_{w}\!\left(
  \mathbf{F}_t \mid \mathbf{F}_0,\sigma_t^2\mathbf{I}
  \right)
  \propto
  \sum_{\mathbf{Z}\in\mathbb{Z}^{N\times 3}}
  \exp\!\left(
  -\frac{\|\mathbf{F}_t-\mathbf{F}_0+\mathbf{Z}\|^2}{2\sigma_t^2}
  \right), \\
  \mathbf{F}_t
  &=
  w(\mathbf{F}_0+\sigma_t\boldsymbol{\epsilon}_F),
  \quad
  \boldsymbol{\epsilon}_F\sim\mathcal{N}(0,\mathbf{I}),
  \end{aligned}
\end{equation}
where the noise schedule $\sigma_t$ obeys the exponential scheduler: $\sigma_0 = 0$ and $\sigma_t = \sigma_1 (\frac{\sigma_T}{\sigma_1})^{\frac{t-1}{T-1}}$, if $t > 0$.

The exact treatment of atom types is model-specific in diffusion-based crystal generative models.
For notation simplicity, we describe the diffusion of atom types using a form similar to the lattice diffusion:
\begin{equation}
\label{eq:diffusion_A}
  q(\mathbf{A}_t \mid \mathbf{A}_0)
  =
  \mathcal{N}\!\left(
  \mathbf{A}_t \mid \sqrt{\bar{\alpha}_t}\mathbf{A}_0,
  (1-\bar{\alpha}_t)\mathbf{I}
  \right),
  \quad
  \mathbf{A}_t
  =
  \sqrt{\bar{\alpha}_t}\mathbf{A}_0
  +
  \sqrt{1-\bar{\alpha}_t}\boldsymbol{\epsilon}_A,
  \quad
  \boldsymbol{\epsilon}_A\sim\mathcal{N}(0,\mathbf{I}).
\end{equation}

According to these forward processes, the corrupted structure $\mathcal{M}_t = (\mathbf{A}_t, \mathbf{F}_t, \mathbf{L}_t)$ can be obtained using the corresponding reparameterized forms shown above.
The denoising model predicts $\hat{\boldsymbol{\epsilon}}_{\theta}^A$, $\hat{\boldsymbol{\epsilon}}_{\theta}^F$, and $\hat{\boldsymbol{\epsilon}}_{\theta}^L$ to regress $\boldsymbol{\epsilon}_{A}$, $\nabla_{\mathbf{F}_t}\log q\left(\mathbf{F}_t \mid \mathbf{F}_0\right)$, and $\boldsymbol{\epsilon}_{L}$.
The denoising model is parameterized by a graph neural network with $K$ message-passing layers.
Let $f^{(K)}(\mathcal{M}_t, t) \in \mathbb{R}^{N \times d_1}$ denote the final atom-wise hidden states.
Based on these hidden states, the model predicts the denoising targets for atom types, fractional coordinates, and lattice variables:
\begin{equation}
  \hat{\boldsymbol{\epsilon}}_{\theta}^A = \varphi_A(f^{(K)}(\mathcal{M}_t, t)),
  \quad 
  \hat{\boldsymbol{\epsilon}}_{\theta}^F = \varphi_F(f^{(K)}(\mathcal{M}_t, t)),
  \quad
  \hat{\boldsymbol{\epsilon}}_{\theta}^L = \varphi_L(\frac{1}{N} \sum_{i}^N f_i^{(K)}(\mathcal{M}_t, t)),
\end{equation}
\begin{equation}
  \mathcal{L}_{A} = \mathbb{E}_{t\sim\mathcal{U}(1,T)}\left[ \left\| \boldsymbol{\epsilon}_{A} - \hat{\boldsymbol{\epsilon}}_{\theta}^A(\mathcal{M}_t,t)
\right\|^2 \right],
\end{equation}
\begin{equation}
  \mathcal{L}_{F} = \mathbb{E}_{t\sim\mathcal{U}(1,T)}[\lambda_t \left\| \nabla_{\mathbf{F}_t}\log q\left(\mathbf{F}_t \mid \mathbf{F}_0\right) - \hat{\boldsymbol{\epsilon}}_{\theta}^F(\mathcal{M}_t,t)  \right\|^2],
\end{equation}
\begin{equation}
  \mathcal{L}_{L} = \mathbb{E}_{t\sim\mathcal{U}(1,T)}\left[ \left\| \boldsymbol{\epsilon}_{L} - \hat{\boldsymbol{\epsilon}}_{\theta}^L(\mathcal{M}_t,t)
\right\|^2 \right],
\end{equation}
where $\varphi_A$, $\varphi_F$, $\varphi_L$ are prediction heads, and $\lambda_t = \mathbb{E}_{\mathbf{F}_t}^{-1}
\left[\left\| \nabla_{\mathbf{F}_t}\log q(\mathbf{F}_t \mid \mathbf{F}_0) \right\|^2\right]$
is approximated via Monte-Carlo sampling.

The total generative loss is:
\begin{equation}
  \mathcal{L}_{\mathrm{diff}} = \lambda_{\mathrm{A}} \mathcal{L}_{\mathrm{A}} + \lambda_{\mathrm{F}} \mathcal{L}_{\mathrm{F}} + \lambda_{\mathrm{L}} \mathcal{L}_{\mathrm{L}}.
\end{equation}

\textbf{Applying CrystalREPA to diffusion models.}
CrystalREPA is applied to the atom-wise hidden states produced during this denoising process.
Given a clean crystal $\mathcal{M}_0=(\mathbf{A}_0,\mathbf{F}_0,\mathbf{L}_0)$, we first construct its corrupted state $\mathcal{M}_t = (\mathbf{A}_t, \mathbf{F}_t, \mathbf{L}_t)$ using the forward diffusion process in Eqs.~(\ref{eq:diffusion_L})--(\ref{eq:diffusion_A}).
The denoising model takes $\mathcal{M}_t$ and the timestep $t$ as input and predicts the corresponding denoising targets for atom types, fractional coordinates, and lattices.
Let $\mathbf{h}_t^{(k)} = f^{(k)}(\mathcal{M}_t,t) \in \mathbb{R}^{N\times d_1}$ denote the atom-wise hidden states from the $k$-th layer of the diffusion model.
The frozen MLIP encoder provides teacher representations $\tilde{\mathbf{z}}=\tilde{f}(\mathcal{M}_0)\in\mathbb{R}^{N\times d_2}$ from the corresponding clean crystal. 
We then project the student representations into the teacher space: $\mathbf{z}=\varphi_p(\mathbf{h}_t^{(k)})\in\mathbb{R}^{N\times d_2}$, where $\varphi_p$ is an MLP.
Finally, we compute the same element-aware contrastive alignment loss $\mathcal{L}_{\mathrm{REPA}}$ as defined in Eqs.~(\ref{eq:s_ij})--(\ref{eq:eance}).

The final training objective for diffusion frameworks is therefore
\begin{equation}
\mathcal{L}
=
\mathcal{L}_{\mathrm{diff}}
+
\lambda_{\mathrm{REPA}}\mathcal{L}_{\mathrm{REPA}}.
\end{equation}

\textbf{Inference.}
Starting from the pure noise state $\mathcal{M}_T = (\mathbf{A}_T, \mathbf{F}_T, \mathbf{L}_T)$, where $\mathbf{A}_T \sim \mathcal{N}(0,\mathbf{I})$, $\mathbf{F}_T \sim \mathcal{U}(0,1)$, $\mathbf{L}_T \sim \mathcal{N}(0,\mathbf{I})$, the reverse updates for lattice variables and atom types are defined by:

\begin{equation}
\label{eq:diffusion_sample_lattice}
  \mathbf{L}_{t-1} = \frac{1}{\sqrt{\alpha_t}}\left(\mathbf{L}_t-\frac{\beta_t}{\sqrt{1-\bar{\alpha}_t}}\hat{\boldsymbol{\epsilon}}_{\theta}^L(\mathcal{M}_t,t)\right) + \sqrt{\frac{\beta_t(1 - \bar{\alpha}_{t-1})}{1 - \bar{\alpha}_{t}}}\boldsymbol{\epsilon}_L,
\end{equation}
\begin{equation}
  \mathbf{A}_{t-1} = \frac{1}{\sqrt{\alpha_t}}\left(\mathbf{A}_t-\frac{\beta_t}{\sqrt{1-\bar{\alpha}_t}}\hat{\boldsymbol{\epsilon}}_{\theta}^A(\mathcal{M}_t,t)\right) + \sqrt{\frac{\beta_t(1 - \bar{\alpha}_{t-1})}{1 - \bar{\alpha}_{t}}} \boldsymbol{\epsilon}_A.
\end{equation}
Fractional coordinates are generated using a two-step predictor-corrector sampler method~\cite{song2021scorebased}:
\begin{equation}
\label{eq:diffusion_sample_coord_predictor}
  \mathbf{F}_{t-\frac{1}{2}} = w\left( \mathbf{F}_t + \left(\sigma_t^2 - \sigma_{t-1}^2\right)\hat{\boldsymbol{\epsilon}}^F_\theta (\mathcal{M}_t, t) + \frac{\sigma_{t-1}\sqrt{\sigma_t^2 - \sigma_{t-1}^2}}{\sigma_t} \boldsymbol{\epsilon}_F \right),
\end{equation}
\begin{equation}
\label{eq:diffusion_sample_coord_corr}
  \mathbf{F}_{t-1} = w\left( \mathbf{F}_{t-\frac{1}{2}} + d_t \hat{\boldsymbol{\epsilon}}^F_\theta (\mathbf{A}_{t-1}, \mathbf{F}_{t-\frac{1}{2}}, \mathbf{L}_{t-1}, t-1) + \sqrt{2d_t}\boldsymbol{\epsilon}_F'\right), \quad d_t = \frac{\gamma \sigma_{t-1}}{\sigma_1},
\end{equation}
where $\boldsymbol{\epsilon}_L$, $\boldsymbol{\epsilon}_A$, $\boldsymbol{\epsilon}_F$, $\boldsymbol{\epsilon}'_F \sim \mathcal{N}(0, \mathbf{I})$, $\gamma$ is the step size of Langevin dynamics.
As in the flow matching case, CrystalREPA does not modify the original diffusion sampling procedure. 
For MatterGen and DiffCSP, we follow the original sampling settings and use $T=1000$ reverse diffusion steps during inference.

\section{Experiment Details}
\label{sec:exp_detail}

\subsection{Training datasets}
\textbf{MP-20.}
MP-20, originally introduced by Xie \textit{et al.}~\cite{xie2022crystal}, is curated from the Materials Project~\cite{jain2013commentary} (Creative Commons Attribution 4.0 International License) and contains 45,231 materials with diverse crystal structures and chemical compositions, spanning 89 elements and unit cells containing 1 to 20 atoms. 
To ensure experimental fidelity, the dataset is restricted to materials originating from the Inorganic Crystal Structure Database (ICSD)~\cite{belsky2002new}, which covers the majority of experimentally reported crystals within this size range. 
It further retains only materials with energy above hull below 0.08 eV/atom and formation energy below 2 eV/atom, and all structures are relaxed using DFT. 
We adopt the original random 60/20/20 split for training, validation, and testing following Xie \textit{et al.}~\cite{xie2022crystal}.

\textbf{Alex-MP-20.}
Alex-MP-20, used in MatterGen~\cite{zeni2025generative}, is a large-scale crystal dataset derived from the Materials Project~\cite{jain2013commentary} and Alexandria~\cite{schmidt2023machine, schmidt2024improving} (Creative Commons Attribution 4.0 International
License) datasets, containing 607,683 structures with up to 20 atoms in the unit cell.
It is constructed by retaining structures with DFT-computed energy above hull below 0.1 eV/atom, while manually excluding structures from well-explored chemical systems, yielding a large and diverse set of stable and metastable crystals.

\subsection{Universal MLIPs}
\label{sec:mlip_details}
We use 10 pretrained universal MLIPs as frozen representation teachers, spanning different model families, pretraining data sources, model sizes, and representation dimensions. 
These teachers include DPA, MACE, ORB, SevenNet, and MatterSim models, as summarized in Tab.~\ref{tab:mlip_details}.
Their pretraining data cover several large-scale atomistic datasets, including 
MPtrj~\cite{deng2023chgnet},
Alexandria~\cite{schmidt2023machine,schmidt2024improving},
sAlex~\cite{barroso2024open},
OMat24~\cite{barroso2024open},
OpenLAM~\cite{openlam-data-v1-web},
COSMOSDataset~\cite{kim2026optimizing}, and 
MatterSim~\cite{yang2024mattersim}.
DPA-3.1-3M-FT denotes the DPA-3.1-3M checkpoint further fine-tuned on MPtrj.

\begin{table}[t]
  \caption{
  Universal MLIP teachers considered in this work.
  We summarize the atom-wise embedding dimension used for representation alignment, the number of model parameters, the approximate scale of the pretraining data, and the main pretraining data sources.
  }
  \label{tab:mlip_details}
  \centering
  \begin{spacing}{1.2}
  \footnotesize
  \begin{tabular}{lcccl}
    \toprule
    Model & Embedding dim. & \#Params & Pretraining scale & Pretraining data sources \\
    \midrule
    DPA-3.1-3M~\cite{zhang2025graphneuralnetworkera} & 128 & 3.27M & 163M & OpenLAM\\ 
    DPA-3.1-3M-FT~\cite{zhang2025graphneuralnetworkera} & 128 & 3.27M & 163M & OpenLAM\\
    DPA-3.1-MPtrj~\cite{zhang2025graphneuralnetworkera} & 128 & 3.27M & 1.58M & MPtrj\\
    MACE-MPA-0~\cite{batatia2025foundation} & 256 & 9.06M & 12M & MPtrj, sAlex \\
    MACE-MP-0~\cite{batatia2025foundation} & 256 & 4.69M & 1.58M & MPtrj \\
    ORB v3~\cite{rhodes2025orbv3atomisticsimulationscale} & 256 & 25.5M & 133M & MPtrj, Alex, OMat24 \\
    ORB v2~\cite{neumann2024orbfastscalableneural} & 256 & 25.2M & 32.1M & MPtrj, Alex \\
    SevenNet-Omni-i12~\cite{kim2026optimizing} & 64 & 54.9M & 243M & COSMOSDataset \\
    SevenNet-l3i5~\cite{park2024scalable} & 64 & 1.17M & 1.58M & MPtrj \\
    MatterSim-v1-5M~\cite{yang2024mattersim} & 256 & 4.55M & 17M & MatterSim \\
    \bottomrule
  \end{tabular}
\end{spacing}
\end{table}

\subsection{Hyperparameters}
\label{sec:hyperparameter}
Tab.~\ref{tab:hyperparameters} summarizes the training hyperparameters used for CrystalREPA across different crystal generative models and training datasets.
All reported experiments are conducted using 80GB NVIDIA A800 GPUs, with the number of GPUs specified in the table.
The CrystalREPA projection head is implemented as a lightweight residual MLP, consisting of one residual block with a linear layer and a SiLU activation function, followed by a linear projection to the MLIP representation dimension.
CrystalREPA only affects training and leaves the sampling procedure unchanged during inference; therefore, all inference settings are identical to those of the corresponding baseline models.

\begin{table}[t]
  \centering
  \tiny
  \caption{Training hyperparameters for CrystalREPA across crystal generative models and datasets.}
  \label{tab:hyperparameters}
  \begin{tabular}{L{1.0cm}L{3.0cm}C{1.0cm}C{1.1cm}C{1.0cm}C{1.1cm}C{1.0cm}C{1.1cm}}
\toprule
  Category & Hyperparameter
  & \multicolumn{2}{c}{CrystalFlow}
  & \multicolumn{2}{c}{MatterGen}
  & \multicolumn{2}{c}{DiffCSP} \\
  \cmidrule(lr){3-4}
  \cmidrule(lr){5-6}
  \cmidrule(lr){7-8}
  & & MP-20 & Alex-MP-20
  & MP-20 & Alex-MP-20
  & MP-20 & Alex-MP-20 \\
  \midrule
  \multirow{12}{*}{Training} 
  & \# Params & 20.9M & 20.9M & 44.6M & 44.6M & 12.3M & 12.3M \\
  & Number of GNN layers & 6 & 6 & 4 & 4 & 6 & 6 \\
  & Atom-wise embedding dim. & 512 & 512 & 512 & 512 & 512 & 512 \\
  & Optimizer & Adam & Adam & Adam & Adam & Adam & Adam \\ 
  & Learning rate scheduling & \multicolumn{2}{c}{ReduceLROnPlateau}  &  \multicolumn{2}{c}{ReduceLROnPlateau}  &  \multicolumn{2}{c}{ReduceLROnPlateau}  \\ 
  & Maximum learning rate & 1e-03 & 1e-03 & 1e-04 & 1e-04 & 1e-04 & 1e-04 \\ 
  & Minimum learning rate & 1e-04 & 1e-04 & 1e-06 & 1e-06 & 1e-05 & 1e-05 \\ 
  & Gradient accumulation steps & 1 & 1 & 4 & 4 & 1 & 1 \\ 
  & Training epochs & 10,000 & 5,000 & 1,900 & 1,200 & 10,000 & 3,000 \\
  & Batch size (per GPU)  & 256 & 128 & 128 & 32 & 256 & 256 \\
  & Number of GPUs & 1 & 4 & 1 & 4 & 1 & 4 \\ 
  & Atom type loss prefactor $\lambda_{A}$ & 10 & 10 & 1 & 1 & 20 & 20 \\ 
  & Coordinate loss prefactor $\lambda_{F}$ & 10 & 10 & 0.1 & 0.1 & 1 & 1 \\ 
  & Lattice loss prefactor $\lambda_{L}$ & 1 & 1 & 1 & 1 & 1 & 1 \\
  \addlinespace[0.2em]
  \hline 
  \addlinespace[0.2em]
  \multirow{4}{*}{CrystalREPA} & Alignment loss & EA-NCE & EA-NCE & EA-NCE & EA-NCE & EA-NCE & EA-NCE \\
  & Alignment loss prefactor $\lambda_{\mathrm{REPA}}$ &  1 & 1 & 1 & 1 & 1 & 1\\
  & Temperature $\tau$ & 0.1 & 0.1 & 0.1 & 0.1 & 0.1 & 0.1\\
  & Alignment layer & 2 & 2 & 2 & 2 & 2 & 2\\
  \bottomrule
  \end{tabular}
\end{table}

\subsection{Formation energy probing}
\label{sec:formation_energy_probing}
We evaluate representation quality by probing formation energy per atom. 
For generative models, we extract atom-wise hidden states from each encoder layer on clean structures at $t=1.0$. 
For each crystal, let $N$ be the number of atoms in the unit cell. 
The $k$-th layer atom-wise representations $\{\mathbf{h}^{(k)}_i\}_{i=1}^{N}$ are aggregated by mean pooling:
\begin{equation}
  \mathbf{h}^{(k)}_{\mathrm{global}}
  =
  \frac{1}{N}\sum_{i=1}^{N}\mathbf{h}^{(k)}_i .
\end{equation}
For MLIP teachers, we apply the same mean-pooling procedure to their atom-wise representations. 
Before fitting the regressor, each feature dimension is standardized using the mean and standard deviation computed on the training split. 
We then train a ridge regression model on the MP-20 training split, select the regularization strength from $\{10^{-3}, 10^{-2}, 10^{-1}, 1, 10, 100\}$ using the validation split, and report the mean absolute error (MAE) on the test split.
The probing results are shown in Fig.~\ref{fig:summary}(a) and Appendix Fig.~\ref{fig:additional_probe}.

\subsection{Definition of CRI, ERI and TRI}
\label{sec:cri}
To quantify the distinguishability of atom-wise representations learned by different teacher MLIPs, we define three resolution metrics: the \emph{Chemical Resolution Index} (CRI), the \emph{Environment Resolution Index} (ERI), and the \emph{Total Resolution Index} (TRI).
These metrics are designed to assess whether the representation space distinguishes chemical identities and reflects similarities in local geometric environments, which we hypothesize to be important for effective generative transfer.

Given a pooled set of $N$ atoms collected from randomly sampled crystals, let $\mathbf{z} \in \mathbb{R}^{N \times d_2}$ be the atom-wise representations extracted from a pretrained universal MLIP.
We first compute the cosine similarity matrix 
\begin{equation}
  \mathbf{S}_{ij} = \frac{\mathbf{z}_i^\top \mathbf{z}_j}{\|\mathbf{z}_i\| \|\mathbf{z}_j\|}.
\end{equation}

\textbf{Chemical Resolution Index (CRI).}
CRI measures how well the representation separates atoms of different elemental species.
We partition all off-diagonal atom pairs into two sets: intra-element pairs, where $\mathbf{A}_i = \mathbf{A}_j$, and inter-element pairs, where $\mathbf{A}_i \neq \mathbf{A}_j$.
We then define
\begin{equation}
  \mathrm{CRI} = \frac{1}{|\mathcal{P}_{\mathrm{intra}}|} \sum_{(i,j)\in \mathcal{P}_{\mathrm{intra}}}\mathbf{S}_{ij} - \frac{1}{|\mathcal{P}_{\mathrm{inter}}|} \sum_{(i,j)\in \mathcal{P}_{\mathrm{inter}}}\mathbf{S}_{ij},
\end{equation}
where $\mathcal{P}_{\mathrm{intra}} = \{(i,j)| i \neq j, \mathbf{A}_i = \mathbf{A}_j\}$ and $\mathcal{P}_{\mathrm{inter}} = \{(i,j)| i \neq j, \mathbf{A}_i \neq \mathbf{A}_j\}$.
A larger CRI indicates that atoms from the same element cluster more tightly in the representation space while atoms from different elements are clearly separated.

\textbf{Environment Resolution Index (ERI).}
ERI measures whether the learned representation is sensitive to local geometric environments.
As a reference for local-environment similarity, we use element-agnostic SOAP descriptors~\cite{bartok2013representing, darby2022compressing}.
Specifically, let $\mathbf{z}^{\mathrm{soap}} \in \mathbb{R}^{N \times d_3}$ denote the corresponding element-agnostic SOAP descriptors computed using the DScribe package~\cite{dscribe, dscribe2}.
We then compute the cosine similarity matrix
\begin{equation}
  \mathbf{S}_{ij}^{\mathrm{soap}} = \frac{(\mathbf{z}_i^{\mathrm{soap}})^\top \mathbf{z}_j^{\mathrm{soap}}}{\|\mathbf{z}_i^{\mathrm{soap}}\| \|\mathbf{z}_j^{\mathrm{soap}}\|}.
\end{equation}
We next compute the mean SOAP similarity over all off-diagonal pairs:
\begin{equation}
  \bar{\mathbf{S}}^{\mathrm{soap}} = \frac{1}{N(N-1)}\sum_{i\neq j}\mathbf{S}_{ij}^{\mathrm{soap}}.
\end{equation}
Using this mean value as a parameter-free threshold, we divide atom pairs into two groups (pairs exactly at the mean threshold are ignored):
\begin{equation}
  \mathcal{P}_{\mathrm{high}} = \{(i,j)|i\neq j,\mathbf{S}_{ij}^{\mathrm{soap}} > \bar{\mathbf{S}}^{\mathrm{soap}}\}, \quad \mathcal{P}_{\mathrm{low}} = \{(i,j)|i \neq j,\mathbf{S}_{ij}^{\mathrm{soap}} < \bar{\mathbf{S}}^{\mathrm{soap}}\}.
\end{equation}
We then define
\begin{equation}
  \mathrm{ERI} = \frac{1}{|\mathcal{P}_{\mathrm{high}}|} \sum_{(i,j)\in \mathcal{P}_{\mathrm{high}}}\mathbf{S}_{ij} - \frac{1}{|\mathcal{P}_{\mathrm{low}}|} \sum_{(i,j)\in \mathcal{P}_{\mathrm{low}}}\mathbf{S}_{ij}.
\end{equation}
A larger ERI indicates that the learned representation assigns higher similarity to atoms with more similar local geometric environments, and lower similarity to atoms in dissimilar environments.

\textbf{Total Resolution Index (TRI).} 
Finally, we combine the above two aspects into a single overall distinguishability score:
\begin{equation}
  \mathrm{TRI} = \mathrm{CRI} + \mathrm{ERI}.
\end{equation}
A larger TRI indicates that the representation is simultaneously discriminative with respect to both chemical species and local structural environment.
We use the unweighted sum because both terms are defined as differences of cosine similarities and therefore lie on comparable scales.
In practice, we pool atom-wise representations from 1,000 sampled crystal structures and compute CRI, ERI, and TRI over all resulting atom pairs.

\subsection{Representation normalization}
\label{sec:rep_norm}

For the diagnostic experiments in Fig.~\ref{fig:rep_norm}, we apply a simple dimension-wise normalization to the frozen MLIP teacher representations before using them as alignment targets.
This normalization is computed once over all atom-wise teacher representations from all structures in the corresponding training split.
Specifically, let $\{\tilde{\mathbf{z}}_i\}_{i=1}^{N_{\mathrm{tot}}}$ denote the flattened set of teacher representations over all atoms in the training split, where $N_{\mathrm{tot}}$ is the total number of atoms.
We compute the dimension-wise mean and standard deviation as
\begin{equation}
  \boldsymbol{\mu}=\frac{1}{N_{\mathrm{tot}}}\sum_{i=1}^{N_{\mathrm{tot}}}\tilde{\mathbf{z}}_i,\qquad \boldsymbol{\sigma}=\sqrt{\frac{1}{N_{\mathrm{tot}}}\sum_{i=1}^{N_{\mathrm{tot}}}(\tilde{\mathbf{z}}_i-\boldsymbol{\mu})^2}.
\end{equation}
The normalized teacher representation is then given by
\begin{equation}
  \tilde{\mathbf{z}}^{\mathrm{norm}}_i=\frac{\tilde{\mathbf{z}}_i-\boldsymbol{\mu}}{\boldsymbol{\sigma}+\epsilon},
\end{equation}
where $\epsilon=10^{-6}$ is used for numerical stability.
The normalized representations are kept fixed and used as alignment targets.

\subsection{Evaluation metrics}
\label{sec:metric_definition}
We follow the MatterGen evaluation protocol~\cite{zeni2025generative} to evaluate generated crystal structures.
For each generative model, we sample 1,024 candidate structures and relax them using MatterSim-v1-1M~\cite{yang2024mattersim}.
Generated structures containing elements not supported by the MatterGen evaluation pipeline are excluded from metric computation.
For each relaxed structure, we compute the energy above the convex hull, denoted as $E_{\mathrm{hull}}$, where the reference convex hull is constructed from 845,997 structures from the Materials Project~\cite{jain2013commentary} and Alexandria~\cite{schmidt2023machine,schmidt2024improving} datasets.
A generated structure is considered metastable if $E_{\mathrm{hull}} \leq 0.1$ eV/atom and stable if $E_{\mathrm{hull}} \leq 0$ eV/atom.
We report \textit{Metastable\%} and \textit{Stable\%} as the fractions of generated structures satisfying these two criteria, respectively.
We also report the average energy above hull, denoted as $\bar{E}_{\mathrm{hull}}$, over all generated structures.

As a proxy for structural fidelity, we report the root mean squared displacement (\textit{RMSD}) between each generated structure before relaxation and its corresponding relaxed structure.
A lower RMSD indicates smaller relaxation-induced structural changes, suggesting that the generated structure is closer to a local equilibrium configuration.

We report \textit{Valid\%} as the fraction of generated structures that pass both structural and compositional validity checks.
For structural validity, we reject structures with interatomic distances smaller than 0.5~\AA{} or unit cell volumes smaller than 0.1~\AA$^3$.
Compositional validity is evaluated using SMACT~\cite{davies2019smact}, which checks whether the generated composition can satisfy charge neutrality.
A generated structure is counted as valid only when both checks are satisfied.

We report \textit{Unique\%} as the fraction of generated structures that do not match any other generated structure in the same batch.
Following MatterGen, uniqueness is computed among structures with the same reduced chemical formula using an ordered-disordered structure matcher.
We report \textit{Novel\%} as the fraction of generated structures that do not match any structure in the reference set used to construct the convex hull, which contains 845,997 entries.
This definition does not necessarily guarantee novelty
with respect to the complete pretraining corpora of all
MLIP teachers.

Following prior work, we further report \textit{MSUN\%} and \textit{SUN\%}, which measure the yield of generated structures that simultaneously satisfy multiple discovery-oriented criteria.
Specifically, MSUN requires a structure to be metastable, unique, and novel, while SUN requires a structure to be stable, unique, and novel.
We report MSUN\% and SUN\% as the corresponding fractions among all generated samples.

\subsection{Conditional generation}
\label{sec:condition_imple}
\subsubsection{CrystalFlow}
For conditional CrystalFlow, we augment the time-dependent vector fields with an external condition $c$, resulting in $\mathbf{v}_\theta^A(\mathcal{M}_t, t, c)$,  $\mathbf{v}_\theta^F(\mathcal{M}_t, t, c)$, and $\mathbf{v}_\theta^L(\mathcal{M}_t, t, c)$.
The original training objective in Eqs.~\eqref{eq:predict}--\eqref{eq:loss_gen} is modified as
\begin{equation}
  \mathcal{L}_{\mathrm{gen}}^{\mathrm{cond}}
  =
  \lambda_A \mathcal{L}_A(\mathbf{v}_\theta^A(\mathcal{M}_t, t, c), \mathbf{u}_t^{A})
  +
  \lambda_F \mathcal{L}_F(\mathbf{v}_\theta^F(\mathcal{M}_t, t, c), \mathbf{u}_t^{F})
  +
  \lambda_L \mathcal{L}_L(\mathbf{v}_\theta^L(\mathcal{M}_t, t, c), \mathbf{u}_t^{L}) .
\end{equation}
In the conditional generation experiments in Appendix~\ref{sec:condition}, we use
energy above hull as the external condition.
During training, $c$ is set to the energy above hull label $c(\mathcal{M})$ of the corresponding clean structure.
All other training hyperparameters are kept the same as those of the vanilla CrystalFlow baseline.

At inference time, we set the target condition to $c_{\mathrm{target}} = 0$ eV/atom and keep CrystalFlow's original Euler sampler unchanged.
The only modification is to replace the velocity field terms in Eqs.~\eqref{eq:Euler_A}--\eqref{eq:Euler_L}, namely
$\mathbf{v}^A_\theta(\mathcal{M}_{t_m},t_m)$,
$\mathbf{v}^F_\theta(\mathcal{M}_{t_m},t_m)$, and
$\mathbf{v}^L_\theta(\mathcal{M}_{t_m},t_m)$,
with their conditional counterparts
$\mathbf{v}^A_\theta(\mathcal{M}_{t_m},t_m,c_{\mathrm{target}})$,
$\mathbf{v}^F_\theta(\mathcal{M}_{t_m},t_m,c_{\mathrm{target}})$, and
$\mathbf{v}^L_\theta(\mathcal{M}_{t_m},t_m,c_{\mathrm{target}})$,
respectively. All other sampling steps follow the original CrystalFlow sampler.

\subsubsection{MatterGen}
For conditional MatterGen, we follow its original property-guided generation strategy based on adapter fine-tuning and classifier-free guidance~\cite{ho2022classifier}.
MatterGen fine-tunes the diffusion score network with condition-dependent adapter modules, while using the same denoising objective as the unconditional model.
In our experiments, we fine-tune the adapter modules for 200 epochs.

In the conditional generation experiments in Appendix~\ref{sec:condition}, we use energy above hull as the external condition.
Following the diffusion convention in Appendix~\ref{sec:crystalrepa_diffusion}, where $\mathcal{M}_0$ denotes the clean structure, each $\mathcal{M}_0$ is paired with its energy above hull label $c(\mathcal{M}_0)$ during adapter fine-tuning.
Following classifier-free guidance, the condition input is randomly replaced by a null condition $c_{\mathrm{null}}$.
Accordingly, the denoising objective is modified as
\begin{equation}
  \begin{aligned}
  \mathcal{L}^{\mathrm{cond}}_{\mathrm{diff}} =
  \lambda_A \mathcal{L}_A
  (\hat{\boldsymbol{\epsilon}}^A_\theta(\mathcal{M}_t,t,\bar{c}), \boldsymbol{\epsilon}_A) 
  & +
  \lambda_F \mathcal{L}_F
  (\hat{\boldsymbol{\epsilon}}^F_\theta(\mathcal{M}_t,t,\bar{c}), \nabla_{\mathbf{F}_t}\log q\left(\mathbf{F}_t \mid \mathbf{F}_0\right)) \\
  & +
  \lambda_L \mathcal{L}_L
  (\hat{\boldsymbol{\epsilon}}^L_\theta(\mathcal{M}_t,t,\bar{c}), \boldsymbol{\epsilon}_L),
  \end{aligned}
\end{equation}
where $\bar c \in \{c(\mathcal{M}_0), c_{\mathrm{null}}\}$ denotes either the
true condition or the null condition.
This allows the same network to provide both conditional predictions
$\hat{\boldsymbol{\epsilon}}_\theta(\mathcal{M}_t,t,c)$ and unconditional
predictions
$\hat{\boldsymbol{\epsilon}}_\theta(\mathcal{M}_t,t,c_{\mathrm{null}})$ for classifier-free guidance.

At inference time, we set the target condition to $c_{\mathrm{target}} = 0$ eV/atom.
The conditional and unconditional denoising predictions are combined as
\begin{equation}
\label{eq:cfg_A}
  \hat{\boldsymbol{\epsilon}}^{A,\gamma}_\theta(\mathcal{M}_t,t,c_{\mathrm{target}}) = \gamma \hat{\boldsymbol{\epsilon}}^A_\theta(\mathcal{M}_t,t,c_{\mathrm{target}}) + (1-\gamma) \hat{\boldsymbol{\epsilon}}^A_\theta(\mathcal{M}_t,t,c_{\mathrm{null}}), 
\end{equation}
\begin{equation}
  \hat{\boldsymbol{\epsilon}}^{F,\gamma}_\theta(\mathcal{M}_t,t,c_{\mathrm{target}}) = \gamma \hat{\boldsymbol{\epsilon}}^F_\theta(\mathcal{M}_t,t,c_{\mathrm{target}}) + (1-\gamma) \hat{\boldsymbol{\epsilon}}^F_\theta(\mathcal{M}_t,t,c_{\mathrm{null}}), 
\end{equation}
\begin{equation}
\label{eq:cfg_L}
  \hat{\boldsymbol{\epsilon}}^{L,\gamma}_\theta(\mathcal{M}_t,t,c_{\mathrm{target}}) = \gamma \hat{\boldsymbol{\epsilon}}^L_\theta(\mathcal{M}_t,t,c_{\mathrm{target}}) + (1-\gamma) \hat{\boldsymbol{\epsilon}}^L_\theta(\mathcal{M}_t,t,c_{\mathrm{null}}),
\end{equation}
where $\gamma$ is the guidance factor.
Following the original MatterGen setting, we set $\gamma=2.0$.

We keep MatterGen's original reverse diffusion sampler unchanged. The only modification is to replace the corresponding denoising prediction terms in Eqs.~\eqref{eq:diffusion_sample_lattice}--\eqref{eq:diffusion_sample_coord_corr},
$\hat{\boldsymbol{\epsilon}}^A_\theta$,
$\hat{\boldsymbol{\epsilon}}^F_\theta$, and
$\hat{\boldsymbol{\epsilon}}^L_\theta$,
with the classifier-free guided predictions defined in Eqs.~\eqref{eq:cfg_A}--\eqref{eq:cfg_L},
$\hat{\boldsymbol{\epsilon}}^{A,\gamma}_\theta$,
$\hat{\boldsymbol{\epsilon}}^{F,\gamma}_\theta$, and
$\hat{\boldsymbol{\epsilon}}^{L,\gamma}_\theta$,
respectively. All other sampling steps follow the original MatterGen sampler.

\subsubsection{DiffCSP}
For conditional DiffCSP, we follow its original energy-guided sampling strategy for property optimization~\cite{zhao2022egsde, bao2023equivariant}, which uses formation energy per atom as the guidance target.
Different from CrystalFlow and MatterGen, DiffCSP does not introduce the target property as an explicit input to the denoising network.
Instead, it trains an additional time-dependent property predictor $E_\psi(\mathcal{M}_t,t)$, which takes an intermediate noisy structure
$\mathcal{M}_t=(\mathbf{A}_t,\mathbf{F}_t,\mathbf{L}_t)$ as input and is supervised by the property label of the corresponding clean structure.

In the conditional generation experiments in Appendix~\ref{sec:condition}, the predictor $E_\psi(\mathcal{M}_t,t)$ is trained to predict the formation energy label $c(\mathcal{M}_0)$ of the corresponding clean structure.
At inference time, we evaluate $E_\psi$ on the current intermediate structure and use its gradients with respect to the current variables to guide the reverse diffusion process toward lower-energy structures.
The sampling process changes from Eqs.~\eqref{eq:diffusion_sample_lattice}--\eqref{eq:diffusion_sample_coord_predictor} to:
\begin{equation}
\label{eq:diffusion_sample_lattice_condition}
  \mathbf{L}_{t-1} = \frac{1}{\sqrt{\alpha_t}}\left(\mathbf{L}_t-\frac{\beta_t}{\sqrt{1-\bar{\alpha}_t}}\hat{\boldsymbol{\epsilon}}_{\theta}^L\right) 
  - s \frac{\beta_t(1-\bar{\alpha}_{t-1})}{1-\bar{\alpha}_t}\nabla_{\mathbf{L}_t} E_\psi
  + \sqrt{\frac{\beta_t(1 - \bar{\alpha}_{t-1})}{1 - \bar{\alpha}_{t}}}\boldsymbol{\epsilon}_L,
\end{equation}
\begin{equation}
  \mathbf{A}_{t-1} = \frac{1}{\sqrt{\alpha_t}}\left(\mathbf{A}_t-\frac{\beta_t}{\sqrt{1-\bar{\alpha}_t}}\hat{\boldsymbol{\epsilon}}_{\theta}^A\right) 
  - s \frac{\beta_t(1-\bar{\alpha}_{t-1})}{1-\bar{\alpha}_t}\nabla_{\mathbf{A}_t} E_\psi
  + \sqrt{\frac{\beta_t(1 - \bar{\alpha}_{t-1})}{1 - \bar{\alpha}_{t}}} \boldsymbol{\epsilon}_A,
\end{equation}
\begin{equation}
\label{eq:diffusion_sample_coord_predictor_condition}
  \mathbf{F}_{t-\frac{1}{2}} = w\left( \mathbf{F}_t + \left(\sigma_t^2 - \sigma_{t-1}^2\right)\hat{\boldsymbol{\epsilon}}^F_\theta 
  - s \frac{\sigma_{t-1}^2(\sigma_t^2-\sigma_{t-1}^2)}{\sigma_t^2}\nabla_{\mathbf{F}_t} E_\psi
  + \frac{\sigma_{t-1}\sqrt{\sigma_t^2 - \sigma_{t-1}^2}}{\sigma_t} \boldsymbol{\epsilon}_F \right),
\end{equation}
where $s$ is the guidance scale.
The subsequent corrector step for fractional coordinates remains unchanged.
Following DiffCSP, we set $s=50$.

\section{Additional Results}
\label{sec:addtional_results}
\subsection{Additional representation probing results}
\label{sec:additional_probe}
\begin{figure}[h]
  \centering
  \includegraphics[width=1.0\textwidth]{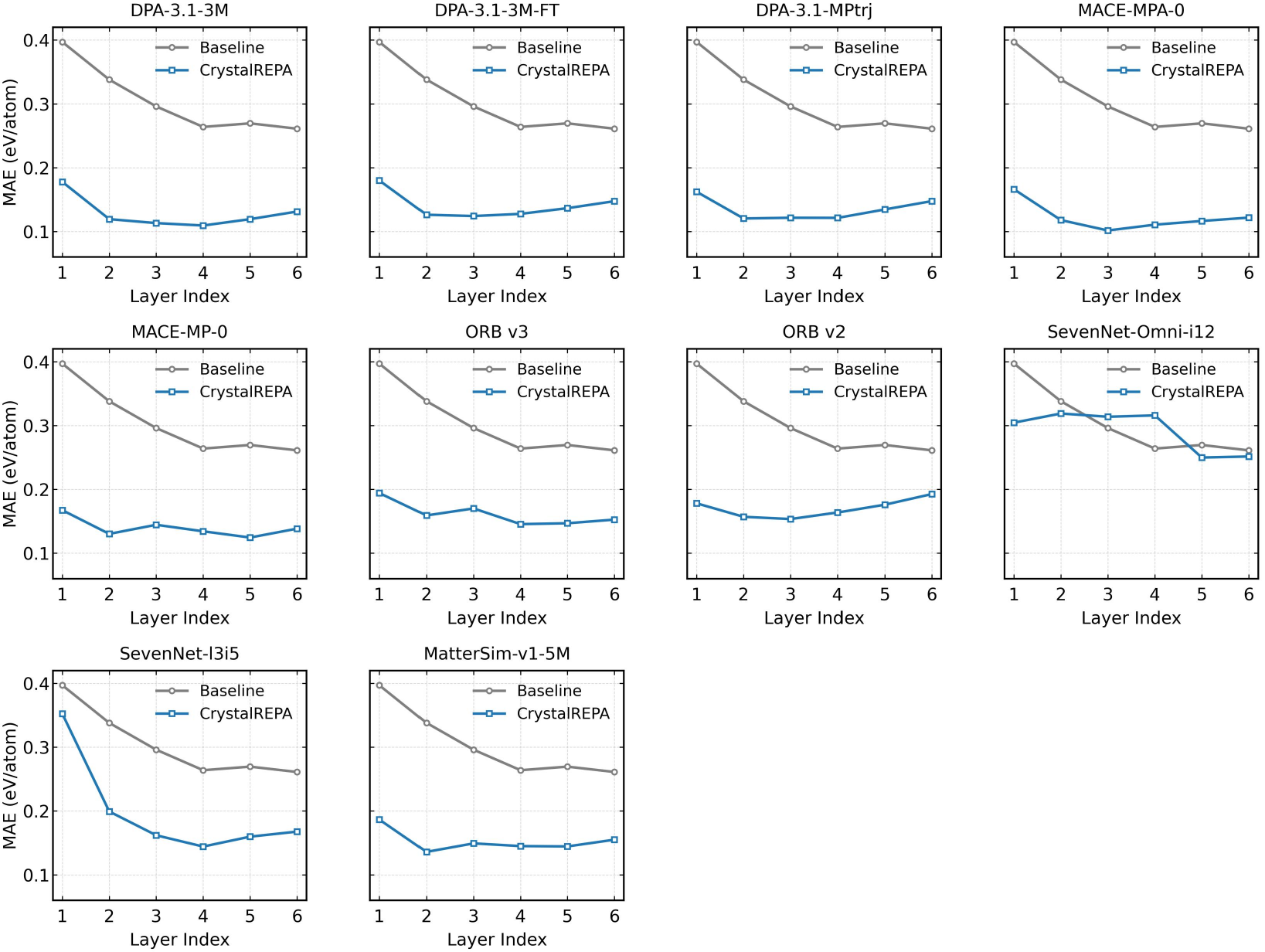}
  \caption{
    Additional formation energy probing results across different MLIP teachers on MP-20.
  }
  \label{fig:additional_probe}
\end{figure}
Fig.~\ref{fig:additional_probe} provides additional formation energy probing results for CrystalREPA trained with different MLIP teachers, following the same protocol as in Appendix~\ref{sec:formation_energy_probing}.
Across most teachers, CrystalREPA reduces the probing MAE compared with vanilla CrystalFlow, especially in early and intermediate layers, suggesting that representation alignment generally improves the formation-energy predictiveness of the generative representations.
One exception is SevenNet-Omni-i12, which shows little probing improvement, consistent with its relatively weak transfer effectiveness in
generation experiments.

\subsection{Additional validation of generated crystals}
\label{app:additional_validation}

\paragraph{DFT-based evaluation.}
To further validate the stability of generated crystals beyond MLIP-based evaluation, we perform DFT calculations on 200 structures generated by vanilla CrystalFlow and 200 structures generated by CrystalFlow with CrystalREPA, using DPA-3.1-3M as the MLIP teacher.
All calculations are performed using VASP 5.4.4 with the projector-augmented-wave (PAW) method and the PBE exchange--correlation functional, following Materials Project-compatible settings, including GGA+U corrections where applicable~\cite{kresse1996efficiency}.
We use a plane-wave cutoff of 520 eV and $\Gamma$-centered k-point meshes with a reciprocal-space density of 64~\AA$^{-3}$.
Atomic positions and cell parameters, including cell shape and volume, are jointly relaxed until the residual forces are below 0.02 eV/\AA.
All relaxations are initialized directly from the generated structures without prior MLIP relaxation.

As shown in Tab.~\ref{tab:dft_validation}, CrystalREPA improves Metastable\% from 50.0 to 64.0 and MSUN\% from 15.0 to 22.5, while reducing $\bar{E}_{\mathrm{hull}}$ from 0.149 to 0.114 eV/atom and RMSD from 0.51 to 0.44.
These results confirm that the improvements in metastability and structural fidelity persist under DFT-based evaluation.

\begin{table}[t]
  \renewcommand{\arraystretch}{1.15}
\centering
\caption{DFT-based evaluation of CrystalFlow and CrystalREPA using DPA-3.1-3M as the MLIP teacher. Each method is evaluated on 200 generated structures.}
\label{tab:dft_validation}
\resizebox{\linewidth}{!}{
\begin{tabular}{lccccccccc}
\toprule
Model & Metastable\%$\uparrow$ & MSUN\%$\uparrow$ & Stable\%$\uparrow$ & SUN\%$\uparrow$ & $\bar{E}_{\mathrm{hull}}\downarrow$ & RMSD$\downarrow$ & Valid\%$\uparrow$ & Unique\%$\uparrow$ & Novel\%$\uparrow$ \\
\midrule
Vanilla CrystalFlow & 50.0 & 15.0 & \textbf{11.0} & \textbf{1.5} & 0.149 & 0.51 & 89.5 & \textbf{100.0} & \textbf{61.5} \\
+ CrystalREPA & \textbf{64.0} & \textbf{22.5} & \textbf{11.0} & \textbf{1.5} & \textbf{0.114} & \textbf{0.44} & \textbf{90.0} & \textbf{100.0} & 55.5 \\
\bottomrule
\end{tabular}
}
\end{table}

\paragraph{Finite-temperature molecular dynamics.}
We further evaluate the dynamical stability of generated crystals through NVT molecular dynamics simulations using the Atomic Simulation Environment (ASE) package~\cite{hjorth2017atomic} with MatterSim-v1.0.0-1M.
Each structure is expanded to a supercell with all lattice dimensions of at least 10~\AA.
Simulations are performed at 100 K using a Langevin thermostat with a timestep of 1 fs.
Each trajectory runs for 25 ps, consisting of 5 ps of equilibration and 20 ps of production dynamics.

A trajectory is considered dynamically stable if it completes successfully and satisfies all the following criteria:
a minimum interatomic distance above 0.75~\AA,
a maximum trajectory RMSD below 1.0~\AA,
an absolute production-stage energy trend below 0.01 eV/atom/ps,
and a mean production-stage temperature within 20 K of the target temperature.
We evaluate 500 generated structures from each of vanilla CrystalFlow and CrystalFlow with CrystalREPA (using DPA-3.1-3M as the MLIP teacher), and report three metrics.
The MD stability rate (MD-SR) is the percentage of structures satisfying all stability criteria.
Max RMSD is the largest root mean square atomic displacement from the starting structure observed during a trajectory.
Max AD is the largest displacement of any individual atom during a trajectory.
Max RMSD and Max AD are reported as medians over the 500 evaluated structures.

As shown in Tab.~\ref{tab:md_validation}, CrystalREPA increases MD-SR from 89.8\% to 93.2\% compared with vanilla CrystalFlow.
It also reduces the median Max RMSD from 0.198 to 0.157~\AA\ and the median Max AD from 0.553 to 0.452~\AA, indicating smaller atomic displacements during finite-temperature dynamics.

\begin{table}[t]
\renewcommand{\arraystretch}{1.15}
\centering
\caption{Finite-temperature MD evaluation of CrystalFlow and CrystalREPA using DPA-3.1-3M as the MLIP teacher.}
\label{tab:md_validation}
\begin{tabular}{lccc}
\toprule
Model & MD-SR (\%)$\uparrow$ & Max RMSD (\AA)$\downarrow$ & Max AD (\AA)$\downarrow$ \\
\midrule
Vanilla CrystalFlow & 89.8 & 0.198 & 0.553 \\
+ CrystalREPA & \textbf{93.2} & \textbf{0.157} & \textbf{0.452} \\
\bottomrule
\end{tabular}
\end{table}

\paragraph{Predicted synthesizability.}
For synthesizability, we adopt the predictor introduced in prior work~\cite{jang2024synthesizability}.
The predictor is trained under a positive-unlabeled learning framework to estimate the synthesizability of an inorganic material from its elemental stoichiometry.
It reports an output probability between 0 and 1, which we use as the synthesizability score, with higher values indicating a greater predicted likelihood of experimental synthesis.
We evaluate the same 1,024 generated structures per model as in Tab.~\ref{tab:result}.
As shown in Tab.~\ref{tab:synthesizability}, CrystalREPA consistently improves the average synthesizability scores across all three generative frameworks.

\begin{table}[t]
\renewcommand{\arraystretch}{1.15}
\centering
\caption{Predicted synthesizability scores of generated crystals using DPA-3.1-3M as the MLIP teacher. Higher scores indicate greater predicted synthesizability.
Values in parentheses denote standard errors over five
independent sampling and evaluation runs.}
\label{tab:synthesizability}
\begin{tabular}{lcc}
\toprule
Model & Vanilla & + CrystalREPA \\
\midrule
CrystalFlow & 0.64(0.00) & \textbf{0.67}(0.00) \\
MatterGen & 0.68(0.00) & \textbf{0.69}(0.00) \\
DiffCSP & 0.62(0.00) & \textbf{0.68}(0.01) \\
\bottomrule
\end{tabular}
\end{table}

\paragraph{Validity-aware MSUN.}
Following the standard MatterGen evaluation protocol~\cite{zeni2025generative}, MSUN\% measures the fraction of generated structures that are simultaneously metastable, unique, and novel, without explicitly requiring structural and compositional validity.
We additionally evaluate Valid-MSUN\%, which requires generated structures to satisfy both the MSUN criteria and the validity checks described in Appendix~\ref{sec:metric_definition}.
For CrystalFlow with DPA-3.1-3M as the MLIP teacher, CrystalREPA improves Valid-MSUN\% from 14.7\% to 17.2\%, consistent with the improvement in MSUN\% from 17.2\% to 20.0\%.
This indicates that the improvement in MSUN\% is not primarily driven by invalid structures.

\subsection{Additional ablation studies}
\label{app:additional_ablations}

\begin{table}[t]
\renewcommand{\arraystretch}{1.15}
\centering
\caption{Comparison of SOAP-based and MLIP-based representation alignment using CrystalFlow. All models are trained for 3,000 epochs, with cosine similarity used as the alignment loss for SOAP- and MLIP-based alignment.}
\label{tab:soap_alignment}
\resizebox{\linewidth}{!}{
\begin{tabular}{llccccccccc}
\toprule
Model & Target Rep. & Metastable\%$\uparrow$ & MSUN\%$\uparrow$ & Stable\%$\uparrow$ & SUN\%$\uparrow$ & $\bar{E}_{\mathrm{hull}}\downarrow$ & RMSD$\downarrow$ & Valid\%$\uparrow$ & Unique\%$\uparrow$ & Novel\%$\uparrow$ \\
\midrule
Vanilla CrystalFlow & -- & 34.8 & 15.5 & 4.1 & 1.3 & 0.224 & 0.64 & 84.5 & \textbf{99.9} & \textbf{78.1} \\
+ CrystalREPA & SOAP & 36.4 & 16.8 & 3.7 & 1.3 & 0.211 & 0.60 & 82.7 & 99.8 & 77.1 \\
+ CrystalREPA & Avg. over 10 MLIPs & \textbf{44.0} & \textbf{18.2} & \textbf{5.1} & \textbf{1.4} & \textbf{0.196} & \textbf{0.57} & \textbf{84.8} & 99.8 & 70.6 \\
\bottomrule
\end{tabular}
}
\end{table}

\paragraph{SOAP alignment control.}
To further investigate the contribution of stability-aware information learned by MLIPs, we compare MLIP-based alignment with alignment using SOAP-based atomic representations~\cite{bartok2013representing}.
Unlike MLIP representations, SOAP descriptors characterize local atomic environments without being learned from energy or force supervision.
Standard SOAP scales poorly with chemical diversity: MP-20 contains 89 elements, which would produce a descriptor with 1,776,796 dimensions and is therefore impractical as an alignment target.
We instead use an element-agnostic SOAP descriptor~\cite{darby2022compressing} with 252 dimensions to encode local geometry and concatenate an 89-dimensional one-hot element encoding, yielding a compact 341-dimensional target representation.

We conduct the comparison using CrystalFlow.
All models are trained for 3,000 epochs, with cosine similarity used as the alignment loss for SOAP- and MLIP-based alignment.
Notably, this SOAP-based representation has a TRI of 1.15, substantially higher than the maximum TRI of 0.62 among the evaluated MLIP representations.

As shown in Tab.~\ref{tab:soap_alignment}, SOAP alignment improves Metastable\%, MSUN\%, and $\bar{E}_{\mathrm{hull}}$ over vanilla CrystalFlow, confirming that geometric and elemental information is useful.
Despite its substantially higher TRI, SOAP alignment still underperforms MLIP-based alignment on the main stability metrics, indicating that geometric and elemental distinguishability, as captured by this SOAP target, is not sufficient to explain the full improvement.

\begin{table}[t]
\renewcommand{\arraystretch}{1.15}
\centering
\caption{Effect of removing the learned formation-energy direction from DPA-3.1-3M representations. All variants are trained for 3,000 epochs using CrystalFlow.}
\label{tab:energy_direction}
\resizebox{\linewidth}{!}{
\begin{tabular}{lccccccccc}
\toprule
Teacher representation & Metastable\%$\uparrow$ & MSUN\%$\uparrow$ & Stable\%$\uparrow$ & SUN\%$\uparrow$ & $\bar{E}_{\mathrm{hull}}\downarrow$ & RMSD$\downarrow$ & Valid\%$\uparrow$ & Unique\%$\uparrow$ & Novel\%$\uparrow$ \\
\midrule
Original DPA-3.1-3M & \textbf{47.9} & \textbf{19.3} & \textbf{6.0} & \textbf{1.6} & \textbf{0.178} & \textbf{0.52} & 85.6 & \textbf{99.8} & 68.0 \\
w/o learned energy direction & 43.1 & 18.0 & 5.0 & 1.3 & 0.199 & 0.63 & 85.0 & \textbf{99.8} & \textbf{71.4} \\
w/o random direction & 47.0 & 19.1 & 5.7 & 1.4 & 0.179 & 0.59 & \textbf{86.0} & 99.6 & 68.6 \\
\bottomrule
\end{tabular}
}
\end{table}

\paragraph{Energy-direction ablation.}
The SOAP comparison shows that MLIP-based alignment achieves stronger stability improvements than alignment with geometric and elemental descriptors.
To further investigate the role of energy-related information, we conduct a diagnostic experiment that removes a learned formation-energy direction from the MLIP teacher representations.

For each crystal $i$, we first mean-pool its atom-wise DPA-3.1-3M representations to obtain
$\bar{\mathbf{z}}_i = \frac{1}{n_i}\sum_{a=1}^{n_i}\mathbf{z}_{i,a}$,
where $n_i$ is the number of atoms in crystal $i$.
Following the probing setup in Fig.~\ref{fig:summary}(a), we fit a ridge regression probe to predict the formation energy per atom.
Specifically, we solve
\begin{equation}
(\hat{\mathbf{w}},\hat{b}) =
\arg\min_{\mathbf{w},b}
\sum_{i=1}^{N}
\left(E_{\mathrm{form},i} -
\mathbf{w}^{\top}\bar{\mathbf{z}}_i - b\right)^2
+ \alpha\|\mathbf{w}\|_2^2.
\end{equation}
The normalized regression coefficient
$\mathbf{u}=\hat{\mathbf{w}}/\|\hat{\mathbf{w}}\|_2$
defines a unit direction associated with formation energy.
For each atom-wise teacher representation, we remove its component along this direction using
\begin{equation}
\mathbf{z}_{i,a}^{\perp}
=
\mathbf{z}_{i,a}
-
(\mathbf{z}_{i,a}^{\top}\mathbf{u})\mathbf{u}.
\end{equation}
We then retrain CrystalREPA on CrystalFlow using the residual representations $\mathbf{z}_{i,a}^{\perp}$ as the alignment targets.
As a dimension-matched control, we repeat the experiment after removing one randomly sampled unit direction.
All variants are trained for 3,000 epochs under the same settings.

As shown in Tab.~\ref{tab:energy_direction}, removing the learned energy direction substantially degrades the main stability metrics, reducing Metastable\% from 47.9 to 43.1 and MSUN\% from 19.3 to 18.0, while increasing $\bar{E}_{\mathrm{hull}}$ from 0.178 to 0.199 eV/atom.
In contrast, removing a randomly sampled direction largely preserves the original performance, achieving 47.0 Metastable\% and 19.1 MSUN\%.
This dimension-matched control suggests that the degradation is specifically associated with removing the learned energy direction, supporting a contribution from energy-related information to downstream generative performance.

\begin{table}[t]
\centering
\renewcommand{\arraystretch}{1.15}
\caption{Effect of the MLIP teacher representation layer on CrystalREPA using CrystalFlow and DPA-3.1-3M. All models are trained for 3,000 epochs. Layer 16 is the final MLIP representation layer.}
\label{tab:teacher_layer}
\resizebox{\linewidth}{!}{
\begin{tabular}{llccccccccc}
\toprule
Model & MLIP Layer & Metastable\%$\uparrow$ & MSUN\%$\uparrow$ & Stable\%$\uparrow$ & SUN\%$\uparrow$ & $\bar{E}_{\mathrm{hull}}\downarrow$ & RMSD$\downarrow$ & Valid\%$\uparrow$ & Unique\%$\uparrow$ & Novel\%$\uparrow$ \\
\midrule
Vanilla CrystalFlow & -- & 34.8(0.3) & 15.5(0.4) & 4.1(0.3) & 1.3(0.1) & 0.224(0.004) & 0.64(0.01) & 84.5(0.4) & \textbf{99.9}(0.0) & \textbf{78.1}(0.6) \\
+ CrystalREPA & 1 & 40.5(0.7) & 18.2(0.3) & 4.3(0.2) & 1.1(0.2) & 0.191(0.004) & 0.63(0.00) & \textbf{85.6}(0.5) & 99.7(0.0) & 73.8(0.8) \\
+ CrystalREPA & 4 & 40.8(0.3) & 17.1(0.3) & 4.4(0.2) & 1.2(0.1) & 0.199(0.001) & \textbf{0.52}(0.01) & 84.3(0.3) & 99.7(0.1) & 73.2(0.6) \\
+ CrystalREPA & 8 & 44.1(0.3) & 18.5(0.2) & 5.1(0.2) & 1.2(0.1) & 0.184(0.005) & 0.56(0.00) & 85.1(0.3) & 99.7(0.0) & 71.0(0.1) \\
+ CrystalREPA & 12 & 45.7(0.6) & 18.8(0.5) & 5.3(0.2) & 1.3(0.1) & \textbf{0.176}(0.005) & 0.58(0.00) & 85.4(0.5) & 99.8(0.1) & 69.9(0.2) \\
+ CrystalREPA & 16 & \textbf{47.9}(0.4) & \textbf{19.3}(0.5) & \textbf{6.0}(0.2) & \textbf{1.6}(0.1) & 0.178(0.005) & \textbf{0.52}(0.01) & \textbf{85.6}(0.2) & 99.8(0.1) & 68.0(0.6) \\
\bottomrule
\end{tabular}
}
\end{table}

\paragraph{MLIP teacher-layer ablation.}
While Fig.~\ref{fig:ablation}(a) examines the alignment layer in the generative model, we further investigate the choice of representation layer in the MLIP teacher.
In our main experiments, we use the final representation layer of the MLIP teacher for alignment.
To examine whether intermediate MLIP representations may be more effective, we conduct a teacher-layer ablation using CrystalFlow as the generative backbone and DPA-3.1-3M as the teacher MLIP.
All models are trained for 3,000 epochs.

As shown in Tab.~\ref{tab:teacher_layer}, all tested teacher representation layers improve Metastable\% and MSUN\% over vanilla CrystalFlow.
Deeper teacher representation layers generally provide stronger stability improvements, with the final representation layer achieving the best performance in Metastable\%, MSUN\%, Stable\%, and SUN\%.
These results support our use of the final MLIP representation layer as a strong default, while also indicating that useful transferable information is present across multiple teacher depths.

\begin{table}[t]
\centering
\renewcommand{\arraystretch}{1.15}
\caption{Comparison of MLIP-only optimization, vanilla DiffCSP, and CrystalREPA using MatterSim-v1-5M. The MLIP-only baseline optimizes randomly initialized structures while retaining the compositions and cell volumes of CrystalREPA-generated structures.}
\label{tab:mlip_optimization}
\resizebox{\linewidth}{!}{
\begin{tabular}{lccccccccc}
\toprule
Model & Metastable\%$\uparrow$ & MSUN\%$\uparrow$ & Stable\%$\uparrow$ & SUN\%$\uparrow$ & $\bar{E}_{\mathrm{hull}}\downarrow$ & RMSD$\downarrow$ & Valid\%$\uparrow$ & Unique\%$\uparrow$ & Novel\%$\uparrow$ \\
\midrule
MLIP-only optimization & 18.1 & 10.1 & 1.8 & 1.6 & 0.439 & 2.04 & 82.5 & \textbf{100.0} & \textbf{90.5} \\
Vanilla DiffCSP & 43.2 & 18.8 & 5.9 & 2.0 & 0.208 & \textbf{0.36} & 82.7 & 98.9 & 72.0 \\
+ CrystalREPA & \textbf{52.6} & \textbf{23.7} & \textbf{7.9} & \textbf{2.9} & \textbf{0.158} & \textbf{0.36} & \textbf{85.4} & 99.4 & 68.0 \\
\bottomrule
\end{tabular}
}
\end{table}

\paragraph{MLIP-only optimization baseline.}
Having examined the role of MLIP representations in CrystalREPA, we further investigate whether direct MLIP-based optimization can achieve comparable generation quality.
To this end, we construct an MLIP-only optimization baseline using MatterSim-v1-5M.
For each structure, we retain the composition and cell volume of a structure generated by CrystalREPA on the DiffCSP framework, using MatterSim-v1-5M as the teacher.
Retaining the cell volume prevents physically unreasonable densities, such as overly compressed cells.
We then randomly initialize the atomic coordinates and lattice shape, rescale the randomized lattice to the retained volume, and jointly optimize the atomic coordinates and lattice parameters using MatterSim-v1-5M.

As shown in Tab.~\ref{tab:mlip_optimization}, direct MLIP optimization from random initialization performs substantially worse than both vanilla DiffCSP and CrystalREPA in Metastable\%, MSUN\%, Stable\%, and SUN\%, while also yielding a higher average energy above hull and substantially larger RMSD.
Although the MLIP-only baseline achieves higher Unique\% and Novel\%, this likely reflects the broad structural diversity introduced by random initialization rather than useful crystal generation, as most resulting structures are not metastable.
These results indicate that MLIP optimization alone cannot replace the structural prior learned by the generative model.

\subsection{Additional representation analysis}
\label{app:additional_representation}

\paragraph{Energy Resolution Index (EnRI).}
To further examine the relationship between thermodynamic stability resolution and transfer effectiveness, we introduce the Energy Resolution Index (EnRI).
EnRI is a structure-level metric that measures how well crystal representations distinguish different thermodynamic stability levels.
We use energy above the hull, $E_{\mathrm{hull}}$, as a proxy for thermodynamic stability.
Since $E_{\mathrm{hull}}$ is defined at the crystal level, we mean-pool the atom-wise MLIP representations to obtain one global representation per crystal.

Given $N$ crystals, let $\mathbf{z}_i \in \mathbb{R}^{d}$ denote the mean-pooled representation of crystal $i$.
We partition the crystals into $K=10$ quantile bins according to $E_{\mathrm{hull},i}$, with approximately equal numbers of crystals per bin.
We then compute the pairwise cosine similarity as
\begin{equation}
S_{ij} =
\frac{\mathbf{z}_i^{\top}\mathbf{z}_j}
{\|\mathbf{z}_i\|_2\|\mathbf{z}_j\|_2}.
\end{equation}

Let $\mathcal{B}_k$ denote the set of crystals in the $k$-th $E_{\mathrm{hull}}$ bin.
The average within-bin similarity is
\begin{equation}
S_{\mathrm{within}} =
\frac{1}{K}\sum_{k=1}^{K}
\frac{1}{|\mathcal{B}_k|(|\mathcal{B}_k|-1)}
\sum_{\substack{i,j\in\mathcal{B}_k\\i\neq j}} S_{ij}.
\end{equation}

The average between-bin similarity is
\begin{equation}
S_{\mathrm{between}} =
\frac{2}{K(K-1)}
\sum_{1\leq k<l\leq K}
\frac{1}{|\mathcal{B}_k||\mathcal{B}_l|}
\sum_{\substack{i\in\mathcal{B}_k\\j\in\mathcal{B}_l}} S_{ij}.
\end{equation}

We define
\begin{equation}
\mathrm{EnRI} =
S_{\mathrm{within}} - S_{\mathrm{between}}.
\end{equation}
A higher EnRI indicates that crystals within similar $E_{\mathrm{hull}}$ ranges have more similar representations than crystals from different $E_{\mathrm{hull}}$ ranges.

\begin{figure}[t]
\centering
\includegraphics[width=0.65\linewidth]{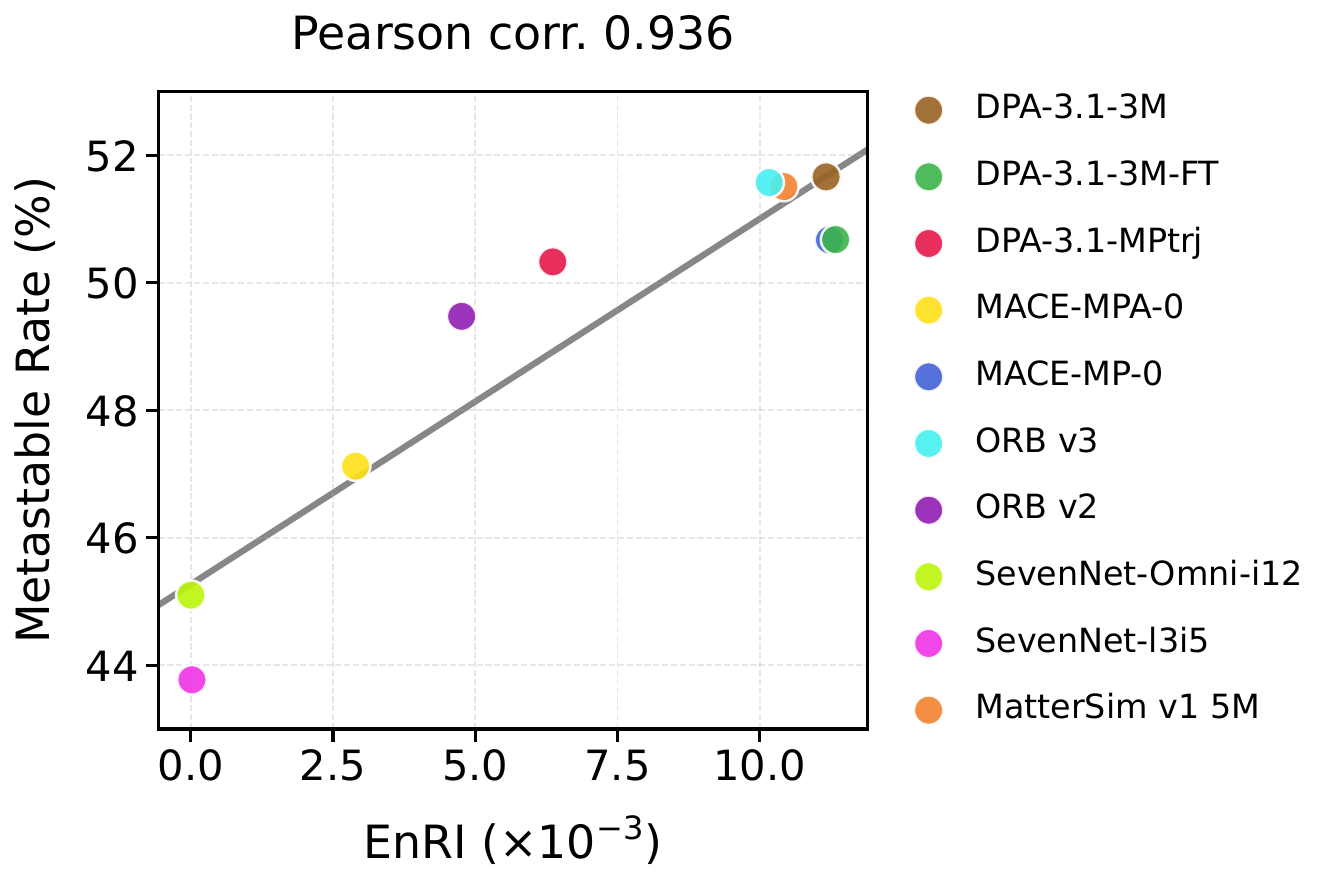}
\caption{Correlation between the Energy Resolution Index (EnRI) and downstream Metastable\% across ten MLIP teachers using CrystalFlow with CrystalREPA. Each point represents an MLIP teacher, and the line indicates a linear fit.}
\label{fig:enri}
\end{figure}

We compute EnRI for all ten MLIP teachers and correlate it with Metastable\% of the corresponding CrystalFlow models trained with CrystalREPA.
As shown in Fig.~\ref{fig:enri}, EnRI exhibits a strong positive Pearson correlation of 0.936 with downstream Metastable\%.
Unlike ERI, which evaluates distinguishability with respect to geometry-derived SOAP similarities, EnRI evaluates distinguishability across thermodynamic stability levels without relying on SOAP.
These results suggest that representations that better distinguish crystals across thermodynamic stability levels tend to yield stronger transfer performance.

\paragraph{Formation-energy probing and generative performance.}
We further examine the relationship between formation-energy probing performance and generative stability across three generative frameworks and their CrystalREPA variants.
Following the linear probing protocol in Appendix~\ref{sec:formation_energy_probing}, we report the formation-energy probing MAE at the second encoder layer, where CrystalREPA applies representation alignment.
For each framework with CrystalREPA, the reported probing MAE and Metastable\% are averaged over ten MLIP teachers.

As shown in Tab.~\ref{tab:probing_correlation}, all three generative frameworks exhibit lower probing MAE and higher Metastable\% after applying CrystalREPA.
Across all 33 evaluated configurations, comprising three baseline models and thirty CrystalREPA variants, probing MAE has a Pearson correlation of $-0.62$ with Metastable\%.
These results further highlight the relationship between stability-related representation quality and downstream generative performance.
Since these configurations differ in architecture, teacher, and training setup, this correlation should be interpreted as an association.

\begin{table}[t]
\centering
\renewcommand{\arraystretch}{1.15}
\caption{Relationship between formation-energy probing performance and generative stability across three frameworks. CrystalREPA results are averaged over ten MLIP teachers.}
\label{tab:probing_correlation}
\resizebox{\linewidth}{!}{
\begin{tabular}{lcccc}
\toprule
Generative framework
& Probe MAE, baseline$\downarrow$
& Probe MAE, +CrystalREPA$\downarrow$
& Metastable\%, baseline$\uparrow$
& Metastable\%, +CrystalREPA$\uparrow$ \\
\midrule
CrystalFlow & 0.338 & 0.158 & 41.2 & 50.5 \\
DiffCSP     & 0.228 & 0.100 & 43.2 & 50.9 \\
MatterGen   & 0.128 & 0.091 & 57.4 & 62.7 \\
\bottomrule
\end{tabular}
}
\end{table}

\subsection{Comparison to conditional generation}
\label{sec:condition}
A natural baseline for improving crystal stability is conditional generation, in which the generator is explicitly conditioned on a stability-related target property.
Here, we compare CrystalREPA against this baseline using energy above hull ($E_{\mathrm{hull}}$) as the conditioning signal.
Implementation details for the conditional baselines are provided in Appendix~\ref{sec:condition_imple}.

As shown in Tab.~\ref{tab:result_condition}, conditional generation improves some stability-related metrics over vanilla generation, but its gains are not consistently better than those of CrystalREPA. 
While conditional generation often improves Stable\%, SUN\% and $\bar{E}_{\mathrm{hull}}$, it consistently reduces Valid\%, indicating a trade-off between target-driven stability optimization and basic structural plausibility. 
By contrast, our proposed CrystalREPA yields more consistent gains on broader quality metrics.
CrystalREPA improves not only post hoc energy-related metrics, but also structural validity, fidelity, and robustness across different generative frameworks.

CrystalREPA also has a practical advantage at inference time. 
It preserves the original sampling procedure and introduces no additional inference overhead.
In contrast, conditional generation typically requires extra inference overhead, such as classifier-free guidance in MatterGen and energy-guidance strategy in DiffCSP.

More importantly, CrystalREPA and conditional generation should be viewed as complementary rather than mutually exclusive. 
CrystalREPA improves the physical quality of the learned generator itself, whereas conditional generation offers explicit controllability toward user-specified objectives. 
Exploring their combination is therefore a promising direction for future goal-directed crystal design.

\begin{table}
  \caption{
    Comparison between CrystalREPA and conditional generation using external energy-related signal.
    For CrystalREPA, “average” reports the mean result across the 10 universal MLIP teachers used in this work, while “best” reports the best result among these teachers for each metric.
    \textbf{Bold} and \underline{underline} denote the best and second-best results, respectively.
  }
  \label{tab:result_condition}
  \centering
  \begin{spacing}{1.2}
  \tiny
  \begin{tabular}
    {L{2.0cm} C{1.0cm}  C{0.9cm} C{0.8cm} C{0.8cm} C{0.9cm} C{0.8cm} C{0.8cm} C{0.8cm} C{0.8cm}}
\toprule
    Model & Metastable\%$\uparrow$ & MSUN\%$\uparrow$ & Stable\%$\uparrow$ & SUN\%$\uparrow$ & $\bar{E}_{\mathrm{hull}}\downarrow$ & RMSD$\downarrow$ & Valid\%$\uparrow$ & Unique\%$\uparrow$ & Novel\%$\uparrow$  \\
    \midrule
    Vanilla CrystalFlow & 41.2 & 17.2 & 5.1 & 1.2 & 0.198 & 0.60 & 84.9 & 99.6 & \textbf{73.1} \\
    Conditional Generation & 40.8 & 16.2 & \underline{6.4} & \underline{1.7} & \underline{0.177} & 0.62 & 81.9 & \underline{99.7} & \underline{72.6} \\
    \rowcolor{gray!5} CrystalREPA (average) & \underline{50.5} & \underline{19.1} & 6.2 & 1.5 & 0.178 & \underline{0.51} & \underline{86.1} & \underline{99.7} & 65.0 \\ 
    \rowcolor{gray!5} CrystalREPA (best) & \textbf{55.0} & \textbf{20.0} & \textbf{7.0} & \textbf{1.8} & \textbf{0.174} & 
    \textbf{0.42} & \textbf{87.5} & \textbf{99.9} & 69.0 \\
  
    \midrule
    \addlinespace[0.5em]
    Vanilla MatterGen & 57.4 & 25.3 & 6.7 & 1.9 & 0.148 & 0.12 & 84.7 & \underline{99.6} & \textbf{64.6}\\
    Conditional Generation & 60.9 & 27.2 & \underline{8.4} & \underline{2.5} & \textbf{0.119} & 0.12 & 83.0 & 99.5 & 63.4 \\
    \rowcolor{gray!5} CrystalREPA (average) & \underline{62.7} & \underline{27.7} & 7.8 & \underline{2.5} & 0.131 & \textbf{0.11} & \underline{85.6} & 99.5 & 61.8 \\
    \rowcolor{gray!5} CrystalREPA (best) & \textbf{66.2} & \textbf{29.2} & \textbf{8.7} & \textbf{2.9} & \textbf{0.119} & \textbf{0.11} & \textbf{86.6} & \textbf{99.7} & \underline{63.5} \\

    \midrule
    \addlinespace[0.5em]
    Vanilla DiffCSP  & 43.2 & 18.8 & 5.9 & 2.0 & 0.208 & 0.36 & 82.7 & 98.9 & \textbf{72.0}\\
    Conditional Generation & \underline{50.9} & 19.7 & 7.1 & \underline{2.7} & 0.169 & \textbf{0.31} & 82.2 & 86.9 & 68.8 \\
    \rowcolor{gray!5} CrystalREPA (average) & \underline{50.9} & \underline{23.4} & \underline{7.2} & 2.6 & \underline{0.166} & 0.37 & \underline{84.3} & \underline{99.2} & 69.3 \\
    \rowcolor{gray!5} CrystalREPA (best) & \textbf{56.5} & \textbf{24.8} & \textbf{8.2} & \textbf{2.9} & \textbf{0.147} & \underline{0.32} & \textbf{85.4} & \textbf{99.5} & \textbf{72.0} \\
    \bottomrule
  \end{tabular}
  \end{spacing}
\end{table}

\subsection{Effect of MLIP teacher input}
\label{sec:teacher_input_ablation}
\begin{table}[t]
  \centering
  \tiny
  \caption{Ablation on the MLIP teacher input for representation alignment.
  Using the clean structure $\mathcal{M}$ as the MLIP input yields better stability-oriented generation quality than using the noisy structure $\mathcal{M}_t$.
  }
  \label{tab:teacher_input}
  \begin{tabular}{lccccccccc}
    \toprule
    MLIP input & Metastable\%$\uparrow$ & MSUN\%$\uparrow$ & Stable\%$\uparrow$ & SUN\%$\uparrow$ & $\bar{E}_{\mathrm{hull}}\downarrow$ & RMSD$\downarrow$ & Valid\%$\uparrow$ & Unique\%$\uparrow$ & Novel\%$\uparrow$  \\
    \midrule
    None & 35.8 & 12.9 & 5.1 & 1.4 & 0.278 & 0.4 & 86.3 & \textbf{99.9} & \textbf{73.8}  \\
    Noisy structure $\mathcal{M}_t$ & 41.6 & 14.6 & 5.6 & 1.8 & 0.237 & \textbf{0.32} & \textbf{87.8} & 99.6 & 69.2 \\
    Clean structure $\mathcal{M}$ & \textbf{50.5} & \textbf{16.5} & \textbf{7.9} & \textbf{2.6} & \textbf{0.204} & 0.34 & 85.8 & 99.6 & 62.0 \\
    \bottomrule
  \end{tabular}
\end{table}
CrystalREPA uses the clean crystal $\mathcal{M}$, rather than the corrupted intermediate structure $\mathcal{M}_t$, as the input to the MLIP teacher. 
The generative encoder computes representations from $\mathcal{M}_t$ to predict the transport or denoising direction toward the clean structure, and the MLIP representation of the clean crystal provides auxiliary supervision consistent with this objective. 
This design is also computationally efficient, since MLIP representations of clean crystals are fixed and can be precomputed offline, whereas using $\mathcal{M}_t$ would require online MLIP forward passes for time-dependent corrupted structures during training.

We conduct this ablation on MP-20 using DiffCSP++~\cite{jiao2024space} as the base generative model and DPA-3.1-3M~\cite{zhang2025graphneuralnetworkera} as the MLIP teacher, with a batch size of 32 and 900,000 training steps.
We compare this design with the natural alternative that feeds $\mathcal{M}_t$ into the MLIP teacher. 
As shown in Tab.~\ref{tab:teacher_input}, using the clean structure $\mathcal{M}$ yields stronger stability-oriented generation quality than using the noisy structure $\mathcal{M}_t$, improving Metastable\%, MSUN\%, Stable\%, and SUN\%, while reducing $\bar{E}_{\mathrm{hull}}$. 
These results support our choice of using the clean structure as the MLIP input and suggest that MLIP representations of clean crystals provide a more effective and efficient supervision signal for stability-oriented representation alignment.

\subsection{Effect of data overlap}
\label{sec:dataoverlap}
\begin{table}[t]
  \centering
  \tiny
  \caption{
  Effect of data overlap between the MLIP teacher pretraining data and the crystal generator training data.
  We construct Alex-MP-clean by removing Alex-MP-20 training structures whose reduced formulas appear in MPtrj.
  DPA-3.1-3M still has substantial reduced formula overlap with Alex-MP-clean through broader pretraining sources such as OMat24, whereas DPA-3.1-MPtrj has no reduced formula overlap with Alex-MP-clean under this filtering criterion.
  All models are trained for 900 epochs.
  }
  \label{tab:data_overlap}
  \begin{tabular}
    {L{1.3cm} L{1.3cm} C{0.4cm} C{0.95cm}  C{0.65cm} C{0.6cm} C{0.5cm} C{0.6cm} C{0.5cm} C{0.5cm} C{0.7cm} C{0.7cm}}
    \toprule
    Model & Target Rep. & Overlap & Metastable\%$\uparrow$ & MSUN\%$\uparrow$ & Stable\%$\uparrow$ & SUN\%$\uparrow$ & $\bar{E}_{\mathrm{hull}}\downarrow$ & RMSD$\downarrow$ & Valid\%$\uparrow$ & Unique\%$\uparrow$ & Novel\%$\uparrow$  \\
    \midrule
    \multicolumn{3}{l}{Vanilla CrystalFlow~\cite{luo2025crystalflow}} & 57.0 & 30.6 & 4.3 & 1.8 & 0.110 & 0.38 & 87.3 & \textbf{100.0} & 70.7 \\
    \rowcolor{gray!5} +~CrystalREPA & DPA-3.1-3M & \checkmark & 58.5 & \textbf{34.0} & 4.3 & \textbf{1.9} & 0.107 & \textbf{0.37} & 88.5 & \textbf{100.0} & \textbf{72.8}\\
    \rowcolor{gray!5} (Ours) & DPA-3.1-MPtrj & $\times$ & \textbf{61.6} & 31.7 & \textbf{5.0} & \textbf{1.9} & \textbf{0.102} & 0.44 & \textbf{88.6} & \textbf{100.0} & 67.2 \\
    \bottomrule
  \end{tabular}
\end{table}
A potential concern is that the improvement from CrystalREPA may be driven by overlap between the MLIP teacher pretraining data and the crystal generator training data, rather than by transferable physical priors.
To examine this issue, we construct a more conservative training split from Alex-MP-20, denoted as Alex-MP-clean.
Specifically, we remove all structures in the Alex-MP-20 training split whose reduced formulas appear in MPtrj. 
This criterion is stricter than exact structure matching, because it removes not only identical structures but also structures with the same reduced formula. 
The resulting Alex-MP-clean split contains 523,090 structures, compared with 607,683 structures in the original Alex-MP-20 training split.

This filtering removes reduced formula overlap with the pretraining data of DPA-3.1-MPtrj. 
In contrast, DPA-3.1-3M is trained on broader data sources beyond MPtrj and therefore still has substantial remaining overlap with Alex-MP-clean.
In particular, 492,438 out of the 523,090 structures in Alex-MP-clean have reduced formulas that appear in the OMat24 training set, which is part of the DPA-3.1-3M pretraining data. 
Therefore, DPA-3.1-3M represents a teacher with substantial remaining reduced formula overlap, while DPA-3.1-MPtrj serves as a non-overlap teacher under the same reduced formula criterion.

As shown in Tab.~\ref{tab:data_overlap}, CrystalREPA remains beneficial on Alex-MP-clean even when using DPA-3.1-MPtrj as the teacher. 
Compared with vanilla CrystalFlow, CrystalREPA with DPA-3.1-MPtrj improves Metastable\% from 57.0 to 61.6 and Stable\% from 4.3 to 5.0, while reducing $\bar{E}_{\mathrm{hull}}$ from 0.110 eV/atom to 0.102 eV/atom. MSUN\% and SUN\% also increase slightly, although RMSD and raw Novel\% are mixed. 
These results indicate that reduced formula overlap with the teacher pretraining data is not necessary for CrystalREPA to improve generation quality.

Moreover, the improvements obtained with DPA-3.1-MPtrj are comparable to those obtained with DPA-3.1-3M, despite the much larger remaining overlap of DPA-3.1-3M with Alex-MP-clean. 
This suggests that training data overlap alone is unlikely to be the primary driver of the observed gains. 
Instead, the results are more consistent with our interpretation that CrystalREPA benefits from transferable stability-aware representations learned by pretrained MLIPs.

\subsection{Computational cost}
\label{sec:computational_cost}
\begin{table}[t]
  \caption{
  Computational cost comparison between baseline crystal generative models and CrystalREPA using DPA-3.1-3M as the teacher MLIP.
  All measurements are conducted on 80GB NVIDIA A800 GPUs.
  Training time follows the settings in Tab.~\ref{tab:hyperparameters}, including the corresponding number of GPUs for each model and dataset.
  Inference time is measured with a unified batch size of 256 for fair comparison.
  }
  \label{tab:cost}
  \centering
  \begin{spacing}{1.2}
  \begin{tabular}{clccc}
    \toprule
    Data & Method & Train / epoch $\downarrow$ & Infer / batch$\downarrow$ & Inference steps \\
    \midrule
\multirow{6}{*}{MP-20}& CrystalFlow & 26.6s & 7.6s & 100\\ 
& +~CrystalREPA & 27.1s & 7.6s & 100\\
    \cmidrule(lr){2-5}
    & MatterGen & 104.4s & 1315.0s & 1000\\
    & +~CrystalREPA & 105.9s & 1315.0s & 1000\\
    \cmidrule(lr){2-5}
    & DiffCSP & 26.6s  &  97.8s & 1000\\
    & +~CrystalREPA & 26.9s & 97.8s & 1000\\
    \midrule
    \multirow{6}{*}{Alex-MP-20}& CrystalFlow & 177.8s & 7.6s & 100\\ 
& +~CrystalREPA & 177.9s & 7.6s & 100\\
    \cmidrule(lr){2-5}
    & MatterGen & 837.8s & 1315.0s & 1000\\
    & +~CrystalREPA & 853.8s & 1315.0s & 1000\\
    \cmidrule(lr){2-5}
    & DiffCSP & 175.5s & 97.8s & 1000\\
    & +~CrystalREPA & 174.4s & 97.8s & 1000\\
    \bottomrule
  \end{tabular}
  \end{spacing}
\end{table}
CrystalREPA introduces only marginal overhead during training and no additional cost at inference time.
In our implementation, the atom-wise teacher representations of clean crystals are precomputed offline and stored with the corresponding training structures.
Therefore, training does not require online forward passes through the pretrained MLIP, and the additional cost comes only from the lightweight projection head and the representation alignment loss.
At inference time, both the teacher encoder and the projection head are removed, so the original sampling procedure is unchanged.

Tab.~\ref{tab:cost} reports the computational cost across three base crystal generative models on MP-20 and Alex-MP-20.
Across the six settings in Tab.~\ref{tab:cost}, CrystalREPA increases the training time per epoch by only 1.0\% on average.
The slightly lower training time observed for DiffCSP on Alex-MP-20 is likely due to normal runtime variation, as the measured difference is small.
These results show that CrystalREPA is a lightweight training-time enhancement while preserving the inference cost of the base generative models.

\section{Broader Impact}
\label{sec:broad_impact}
This work focuses on a methodological improvement for crystal generative modeling. 
Its potential societal impact is expected to be indirect, mainly through downstream use in computational materials discovery. 
By improving the quality of generated crystal candidates, CrystalREPA may provide better initial candidates for downstream computational screening. 
However, generated structures should undergo further computational or experimental validation before practical use.
We do not identify immediate negative societal impacts specific to the proposed method.

\newpage
\section*{NeurIPS Paper Checklist}

\begin{enumerate}

\item {\bf Claims}
    \item[] Question: Do the main claims made in the abstract and introduction accurately reflect the paper's contributions and scope?
    \item[] Answer: \answerYes{} 
    \item[] Justification: The claims made in the abstract and introduction are empirically verified in Section~\ref{sec:experiments}.

\item {\bf Limitations}
    \item[] Question: Does the paper discuss the limitations of the work performed by the authors?
    \item[] Answer: \answerYes{} 
    \item[] Justification: We discuss our limitations in Section~\ref{sec:conclusion}.

\item {\bf Theory assumptions and proofs}
    \item[] Question: For each theoretical result, does the paper provide the full set of assumptions and a complete (and correct) proof?
    \item[] Answer: \answerNA{} 
    \item[] Justification: This work does not present theoretical results or formal proofs.

    \item {\bf Experimental result reproducibility}
    \item[] Question: Does the paper fully disclose all the information needed to reproduce the main experimental results of the paper to the extent that it affects the main claims and/or conclusions of the paper (regardless of whether the code and data are provided or not)?
    \item[] Answer: \answerYes{} 
    \item[] Justification: We disclose all information needed for reproduction in Appendix~\ref{sec:exp_detail}.

\item {\bf Open access to data and code}
    \item[] Question: Does the paper provide open access to the data and code, with sufficient instructions to faithfully reproduce the main experimental results, as described in supplemental material?
    \item[] Answer: \answerYes{} 
    \item[] Justification: We provide publicly available code and instructions for reproducing the main CrystalREPA experiments at \url{https://github.com/Chengqian-Zhang/CrystalREPA}.

\item {\bf Experimental setting/details}
    \item[] Question: Does the paper specify all the training and test details (e.g., data splits, hyperparameters, how they were chosen, type of optimizer) necessary to understand the results?
    \item[] Answer: \answerYes{} 
    \item[] Justification: We provide experimental details in Appendices~\ref{sec:exp_detail} and~\ref{sec:addtional_results}.

\item {\bf Experiment statistical significance}
    \item[] Question: Does the paper report error bars suitably and correctly defined or other appropriate information about the statistical significance of the experiments?
    \item[] Answer: \answerYes{} 
    \item[] Justification: We report standard errors over five independent sampling and evaluation runs for the generation results in Table~\ref{tab:result}, Table~\ref{tab:result_alex_mp} and Figure~\ref{fig:ablation}.

\item {\bf Experiments compute resources}
    \item[] Question: For each experiment, does the paper provide sufficient information on the computer resources (type of compute workers, memory, time of execution) needed to reproduce the experiments?
    \item[] Answer: \answerYes{} 
    \item[] Justification: We provide the information on the computer resources in Appendix~\ref{sec:hyperparameter} and Appendix~\ref{sec:computational_cost}.
    
\item {\bf Code of ethics}
    \item[] Question: Does the research conducted in the paper conform, in every respect, with the NeurIPS Code of Ethics \url{https://neurips.cc/public/EthicsGuidelines}?
    \item[] Answer: \answerYes{} 
    \item[] Justification: We have read and obey the NeurIPS Code of Ethics.

\item {\bf Broader impacts}
    \item[] Question: Does the paper discuss both potential positive societal impacts and negative societal impacts of the work performed?
    \item[] Answer: \answerYes{} 
    \item[] Justification: We discuss the broader impact of this work in the Appendix~\ref{sec:broad_impact}.
    
\item {\bf Safeguards}
    \item[] Question: Does the paper describe safeguards that have been put in place for responsible release of data or models that have a high risk for misuse (e.g., pre-trained language models, image generators, or scraped datasets)?
    \item[] Answer: \answerNA{} 
    \item[] Justification: This work does not release high-risk data or models.
    It focuses on crystal generation for materials discovery using existing benchmark datasets.

\item {\bf Licenses for existing assets}
    \item[] Question: Are the creators or original owners of assets (e.g., code, data, models), used in the paper, properly credited and are the license and terms of use explicitly mentioned and properly respected?
    \item[] Answer: \answerYes{} 
    \item[] Justification: We cite the original papers for the datasets, models, and evaluation tools used in this work.

\item {\bf New assets}
    \item[] Question: Are new assets introduced in the paper well documented and is the documentation provided alongside the assets?
    \item[] Answer: \answerYes{} 
    \item[] Justification: We provide publicly available code and documentation for reproducing the main CrystalREPA experiments at \url{https://github.com/Chengqian-Zhang/CrystalREPA}.

\item {\bf Crowdsourcing and research with human subjects}
    \item[] Question: For crowdsourcing experiments and research with human subjects, does the paper include the full text of instructions given to participants and screenshots, if applicable, as well as details about compensation (if any)? 
    \item[] Answer: \answerNA{} 
    \item[] Justification: This work does not involve crowdsourcing experiments or research with human subjects.

\item {\bf Institutional review board (IRB) approvals or equivalent for research with human subjects}
    \item[] Question: Does the paper describe potential risks incurred by study participants, whether such risks were disclosed to the subjects, and whether Institutional Review Board (IRB) approvals (or an equivalent approval/review based on the requirements of your country or institution) were obtained?
    \item[] Answer: \answerNA{} 
    \item[] Justification: This work does not involve crowdsourcing experiments or research with human subjects.

\item {\bf Declaration of LLM usage}
    \item[] Question: Does the paper describe the usage of LLMs if it is an important, original, or non-standard component of the core methods in this research? Note that if the LLM is used only for writing, editing, or formatting purposes and does \emph{not} impact the core methodology, scientific rigor, or originality of the research, declaration is not required.
    \item[] Answer: \answerNA{} 
    \item[] Justification: LLMs were used only for writing assistance and did not affect the core methodology, experiments, or evaluation.

\end{enumerate}

\end{document}